\documentclass[12pt]{article}\usepackage[colorlinks=true,citecolor=darkgreen,linkcolor=black,urlcolor=purple]{hyperref}
\usepackage{xcolor}\usepackage{color}\definecolor{darkgreen}{rgb}{0,0.5,0}%\usepackage[hyperfootnotes=false]{hyperref}
\usepackage{epsfig}
\usepackage{float}
\usepackage{empheq}
\usepackage{bbold}
\usepackage[utf8]{inputenc}
\usepackage{amsmath}

\usepackage{caption}

\usepackage{amsmath}
\usepackage{amssymb}
\usepackage{graphicx}
\newlength{\abstractwidth}
\renewcommand{\thefootnote}{\fnsymbol{footnote}}
\renewcommand{\thanks}[1]{\footnote{#1}} % Use this for footnotes
\newcommand{\starttext}{
\setcounter{footnote}{0}
\renewcommand{\thefootnote}{\arabic{footnote}}}

\newcommand{\be}{\begin{equation}}
\newcommand{\bea}{\begin{eqnarray}}
\newcommand{\eea}{\end{eqnarray}}
\newcommand{\beq}{\begin{equation}}
\newcommand{\ee}{\end{equation}}

\newcommand*\widefbox[1]{\fbox{\hspace{2em}#1\hspace{2em}}}

\def\eq{&=&}

\def\la{\langle}
\def\ra{\rangle}
\def\simleq{\; \raise0.3ex\hbox{$<$\kern-0.75em
\raise-1.1ex\hbox{$\sim$}}\; }
\def\simgeq{\; \raise0.3ex\hbox{$>$\kern-0.75em
\raise-1.1ex\hbox{$\sim$}}\; }

\def\bi{\begin{itemize}}
\def\ei{\end{itemize}}
\def\S{Schwarzschild}
\def\sc{\setcounter{equation}{0}}
\def\dof{degrees of freedom }
\def\CA{{\cal{A}}}
\def\CB{{\cal{B}}}

\def\Tr{\rm Tr \it}
\def\bsub{ \begin{subequations}
\begin{empheq}[box=\widefbox]{align}  }
\def\esub{ \end{empheq}
\end{subequations}}

\def\1{\(  \mathbb{1} \)}

  \def\kl{k-local}

  \def\bn{\bigskip \noindent}

 \def\bm{\begin{bmatrix}}
 \def\em{\end{bmatrix}}

    \def\bts{bit-threads}
       \def\bt{bit-thread}

\makeatletter
\g@addto@macro\normalsize{%
  \setlength\abovedisplayskip{10pt}
  \setlength\belowdisplayskip{20pt}
  \setlength\abovedisplayshortskip{10pt}
  \setlength\belowdisplayshortskip{20pt}
}
\makeatother

\usepackage{color}

\begin{document}

%%%%%%%%%%%%%%%%
%%%%%%%%%%%%%%%%

%%%%%%%%%%%%%%%%%%%
%%%%%%%%%%%%%%%%%%%%
  
\begin{titlepage}

\rightline{}
\bigskip
\bigskip\bigskip\bigskip\bigskip
\bigskip

\centerline{\Large \bf { Entanglement in De Sitter Space }} 

\bn

\bigskip
\begin{center}
\bf Edgar Shaghoulian$^1$ and     Leonard Susskind$^{2,3}$  \rm

\bigskip
$^1$David Rittenhouse Laboratory, University of Pennsylvania, Philadelphia, PA 19104, USA

$^2$SITP, Stanford University, Stanford, CA 94305, USA \vskip0.4em
$^3$Google, Mountain View, CA 94043, USA\vskip0.4em

%\vspace{1cm}
\end{center}

\bn

\begin{abstract}
This paper expands on two  recent  proposals, \cite{Susskind:2021dfc}\cite{Susskind:2021esx} and \cite{Shaghoulian:2021cef},  for  generalizing the  Ryu-Takayanagi and Hubeny-Rangamani-Takayanagi formulas  to de Sitter space. The proposals (called the monolayer and bilayer proposals) are similar; both    replace the boundary of AdS by the boundaries of static-patches--in other words event horizons. After stating the rules for each,  we  apply them to a number of cases and show that they yield results expected on other grounds.

The monolayer and bilayer proposals often give the same results, but in one particular situation they disagree. To definitively decide between them we need to understand more about the nature of the thermodynamic limit of holographic systems.

\end{abstract}

\end{titlepage}

\starttext \baselineskip=17.63pt \setcounter{footnote}{0}

%\Large

\tableofcontents
%\color{blue}

\sc
\section{Preliminaries}\label{Prelim}
We will begin with some preliminaries concerning the holographic principle for static patches in de Sitter space. For more detail the reader is referred to \cite{Susskind:2021omt}.

\subsection{The Static Patch} \label{SP}
\begin{figure}[H]
\begin{center}
\includegraphics[scale=.3]{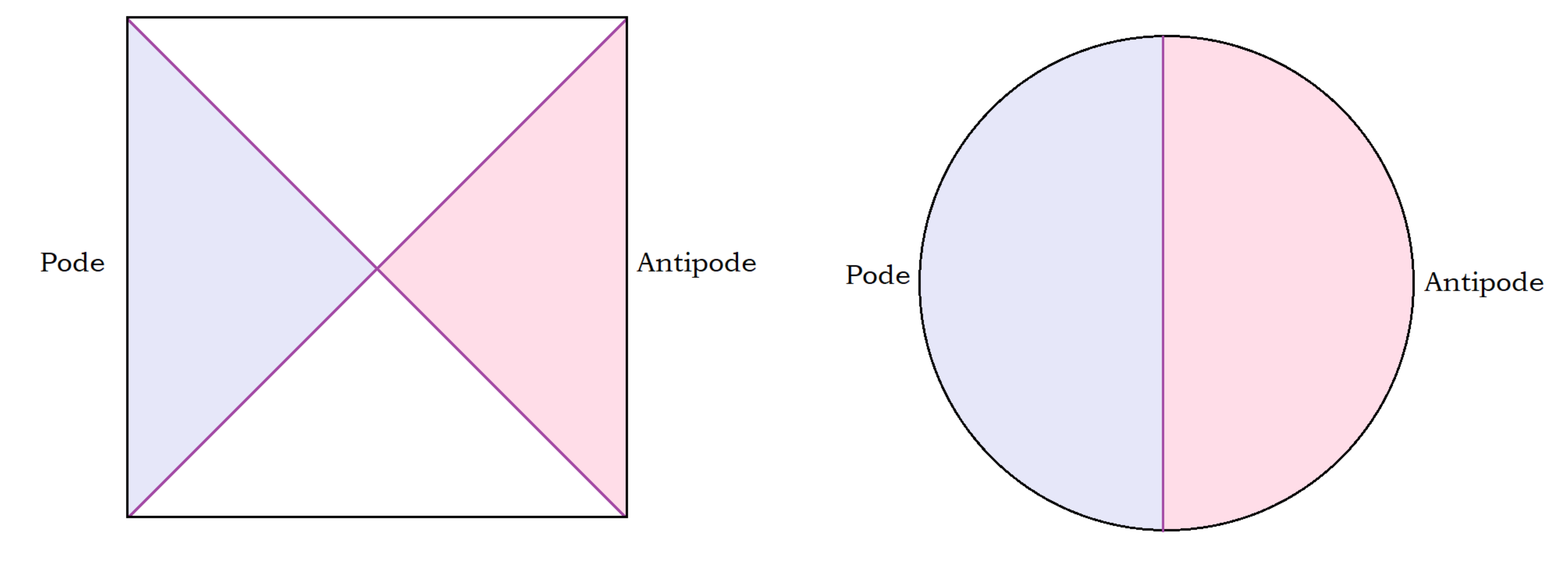}
\caption{The left panel shows the Penrose diagram for de Sitter space with the pode and antipode static patches shown shaded. The right panel shows a time-symmetric spatial slice which forms a $(D-1)$-sphere. The horizons in both panels are colored purple. In the right panel the horizon is the bifurcate horizon which forms a $(D-2)$-sphere.}
\label{penrose-slice}
\end{center}
\end{figure}

The metric of de Sitter space in static coordinates is given by
\bea
ds^2 \eq -f(r)dt^2 +f(r)^{-1}dr^2 +r^2 d\Omega^2, \cr \cr
f(r) \eq 1-\frac{r^2}{R^2}.
\label{metric}
\eea
The static patch is the region for which $f(r)>0,$ in other words where 
$r<R.$ 
Static patches come in complementary pairs as shown in figure \ref{penrose-slice}.
The centers of the static patches $r=0$ are called the pode and the antipode. They  represents the location of  nominal observers.

\subsection{Boundary of a Static patch}
The boundary of a static patch  is its cosmic horizon. In the holographic description of de Sitter space the stretched horizon plays the role of the hologram, i.e., the location of the holographic degrees of freedom. The static patches will be referred to as the interior regions. The other regions of the Penrose diagram (shown as white) which lie beyond the horizons are the exterior regions.

\subsection{Generalized Horizon}
\begin{figure}[H]
\begin{center}
\includegraphics[scale=.3]{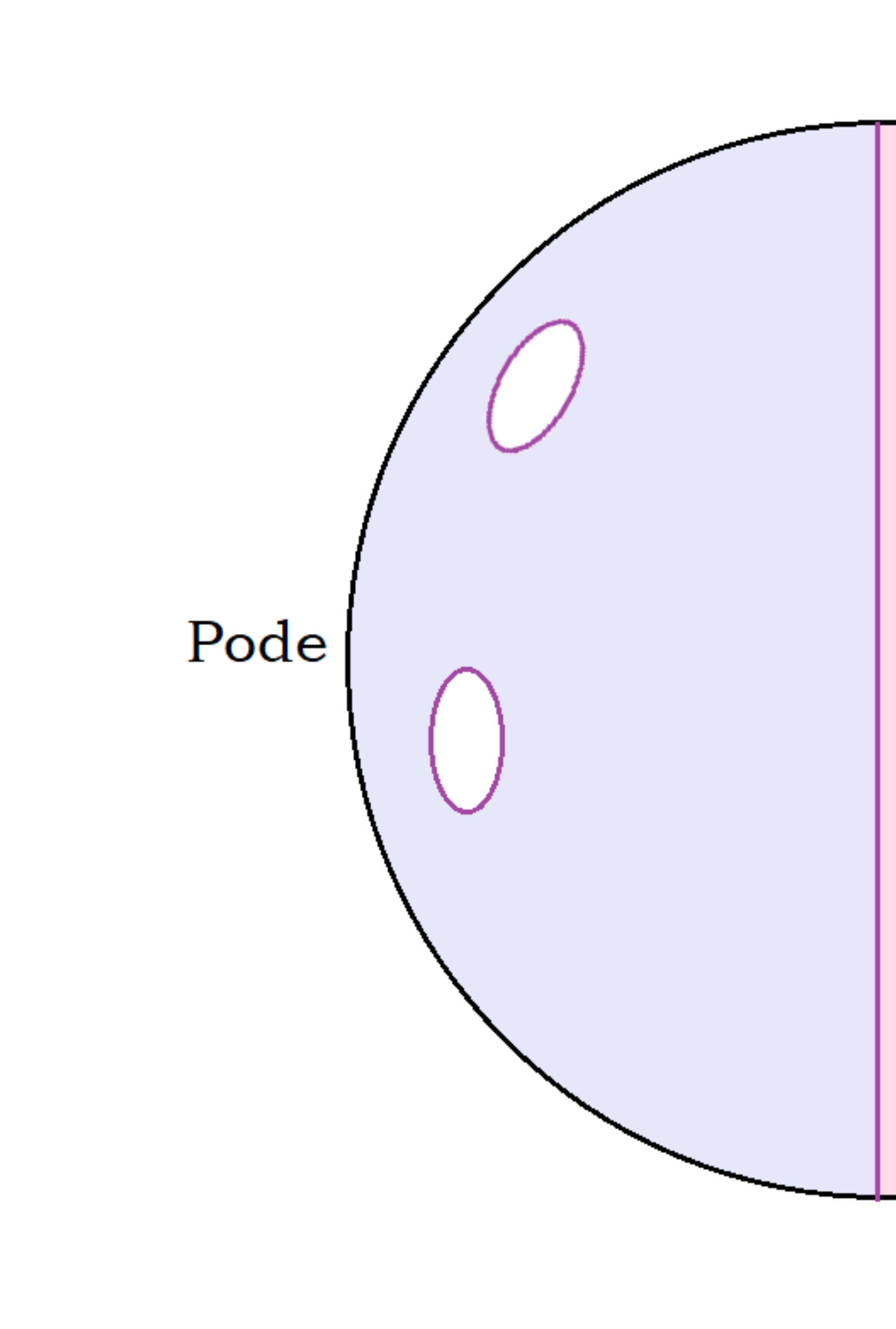}
\caption{The generalized horizon of the pode patch contains the
cosmic horizon as the largest component, and also horizons of black holes that reside in the patch.}
\label{gen-hor}
\end{center}
\end{figure}

More general deformations of pure de Sitter space may contain matter that deforms the geometry. In particular the static patches, which may no longer be static, may contain black holes. The horizons of those black holes define part of the boundary of the observable universe for an observer in the static patch. We will therefore define a \it generalized horizon \rm  which includes not only the cosmic horizon but also all the black hole horizons in the static patch. An example is shown in figure \ref{gen-hor}.

Generalized horizons can change with time, for example as a black hole falls through the cosmic horizon. We will discuss this further in section \ref{asymbh}.

\sc
\section{An Entanglement Proposal}

The proposal of \cite{Susskind:2021dfc}\cite{Susskind:2021esx} and that of  \cite{Shaghoulian:2021cef}, for a geometric  theory of entanglement in de Sitter space, are based on the framework of static-patch holography. They resemble the Ryu-Takayanagi  \cite{Ryu:2006bv} theory except that  the boundaries of  anti de Sitter space are replaced by the boundaries of  static-patches--in other words horizons. (Previous work on static patch holography includes \cite{Alishahiha:2004md}\cite{Banks:2006rx}\cite{Anninos:2011af}\cite{Anninos:2017hhn}\cite{Dong:2018cuv}\cite{Anninos:2018svg}\cite{Coleman:2021nor} and previous proposals for anchoring to the cosmic horizon include \cite{Sanches:2016sxy}\cite{Nomura:2017fyh}.)
We'll refer  to the original proposal \cite{Susskind:2021dfc}\cite{Susskind:2021esx} as the monolayer proposal, and \cite{Shaghoulian:2021cef} as the bilayer proposal. The two are similar and often give the same results, but they are not the same. As we will see, there are situations where the monolayer and bilayer theories give different results for entanglement entropy.  We will begin with the monolayer proposal.

\subsection{Geometry of a Spatial Slice}

\begin{figure}[H]
\begin{center}
\includegraphics[scale=.3]{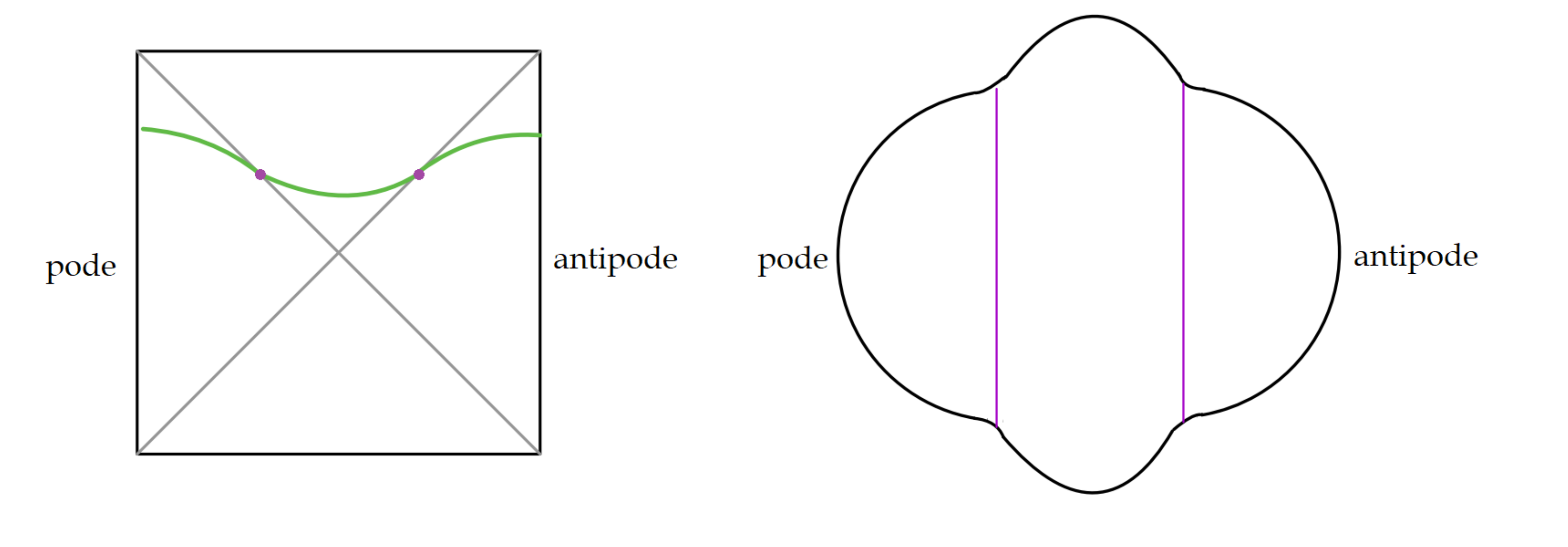}
\caption{A space-like slice through de Sitter space. In the static patches the slice follows constant time ($t$) surfaces until it gets near the horizons and then bends to cross the horizons. The static patches appear as hemispheres of radius $R.$ The exterior region between the horizons bulges to a larger size due to the growth of de Sitter space in the exterior as one moves away from the horizons.}
\label{pode}
\end{center}
\end{figure}

The geometry of spatial slices of de Sitter space will play a prominent role in what follows. The right panel of figure \ref{penrose-slice} shows a special time-symmetric slice that passes through the bifurcate horizon. We will be interested in more general slices which don't pass through the bifurcate horizon.
In figure \ref{pode} the left panel shows the Penrose diagram for de Sitter space along with a spatial slice marked in green. The spatial slice crosses the horizons  at the purple dots.
The bulge between the horizons, which is a kind of Python's Lunch \cite{Brown:2019rox}, has its origin in the inflating property of de Sitter space. As one moves into the region between the horizons the local $(D-2)$-spheres grow and the result is the bulge. Note that in the region between the horizons the areas of the  local  $(D-2)$-spheres have the minimum value at the horizons. That fact plays an important role in the determination of entanglement entropy.

\subsection{Bit Threads}
\begin{figure}[H]
\begin{center}
\includegraphics[scale=.4]{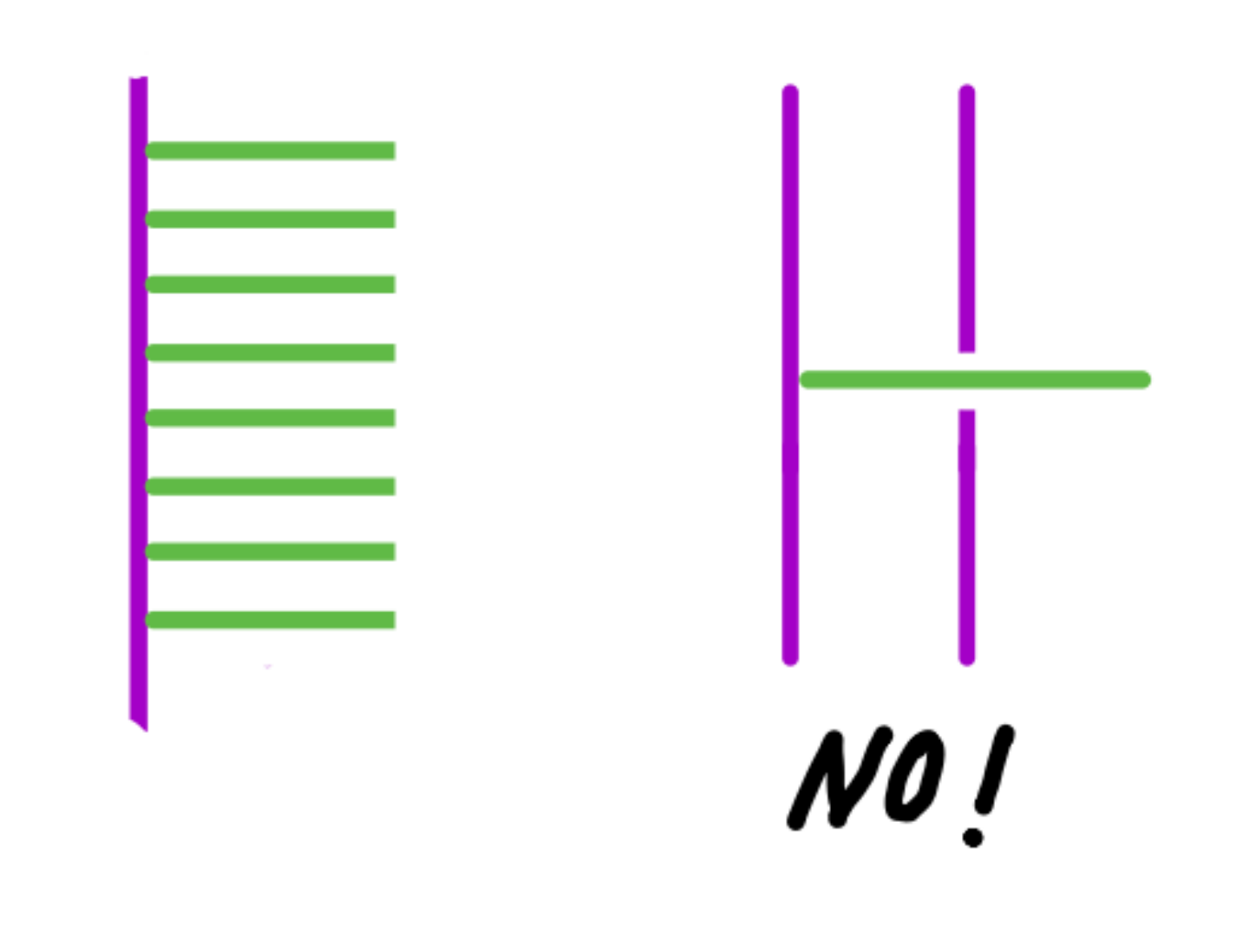}
\caption{Bit-threads emitted from a horizon. The maximum area-density \bts \ is $1/4G.$  Bit-threads may not cross a horizon.}
\label{yesno}
\end{center}
\end{figure}

We will use the 
 bit-thread formulation of Ryu-Takayanagi  \cite{Ryu:2006bv} due to  Freedman and Headrick \cite{Freedman:2016zud}. It is equivalent to the minimal area formulation of  Ryu-Takayanagi, but has the advantage of being  particularly easy to visualize. The horizon surfaces in this formulation are sources of bit threads which emanate from the horizons as illustrated in figure \ref{yesno}. The \bts \ have Planckian thickness and are impenetrable so that their area-density can never exceed $1/4G.$

The relation between bit-threads and entanglement entropy will be explained shortly.

\subsection{Entanglement Entropy}

\begin{figure}[H]
\begin{center}
\includegraphics[scale=.3]{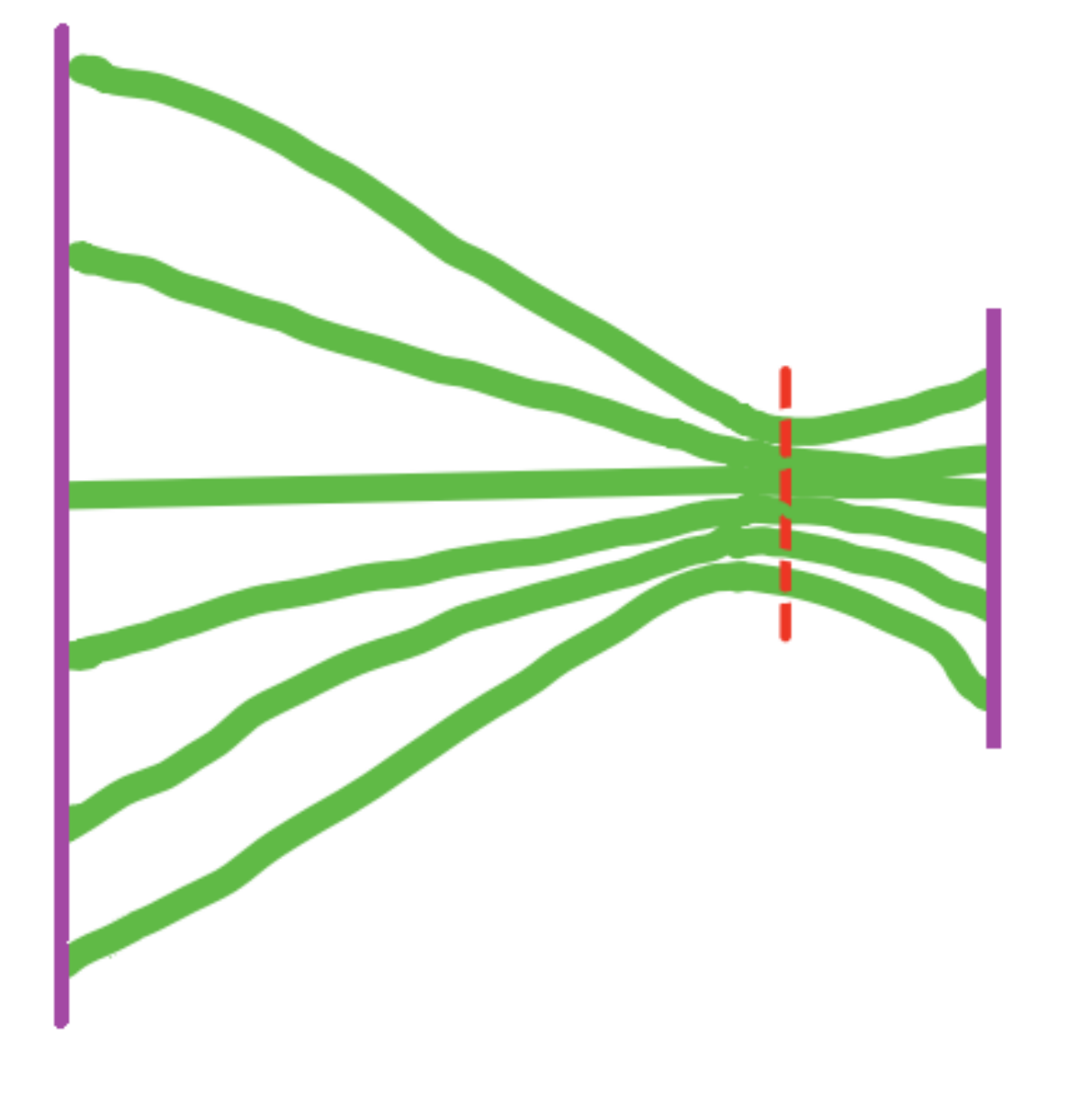}
\caption{The Ryu-Takayanagi minimal area surface defines a bottleneck for \bts.}
\label{bitthread}    
\end{center}
\end{figure}

We begin by dividing the horizon degrees of freedom into two subsets which we will call $\CA$ and $\CB.$ We then draw bit-threads connecting the two subsets. The bit-threads have Planckian thickness and are impenetrable. The entanglement entropy between $\CA$ and $\CB$ is the maximum number of bit-threads that can connect the two subsets. The maximum number is determined by the smallest area of any surface lying between the two subsets. This is illustrated in figure \ref{bitthread}.

One point to emphasize is that the locations of the \bts \ on the boundary (or horizon in the de Sitter case) have no meaning. There may be many ways of moving the locations of the \bts \ that saturate the maximum number. All are equally valid and the \bt \ rules do not distinguish them.

\sc
\section{Pure dS}

\begin{figure}[H]
\begin{center}
\includegraphics[scale=.3]{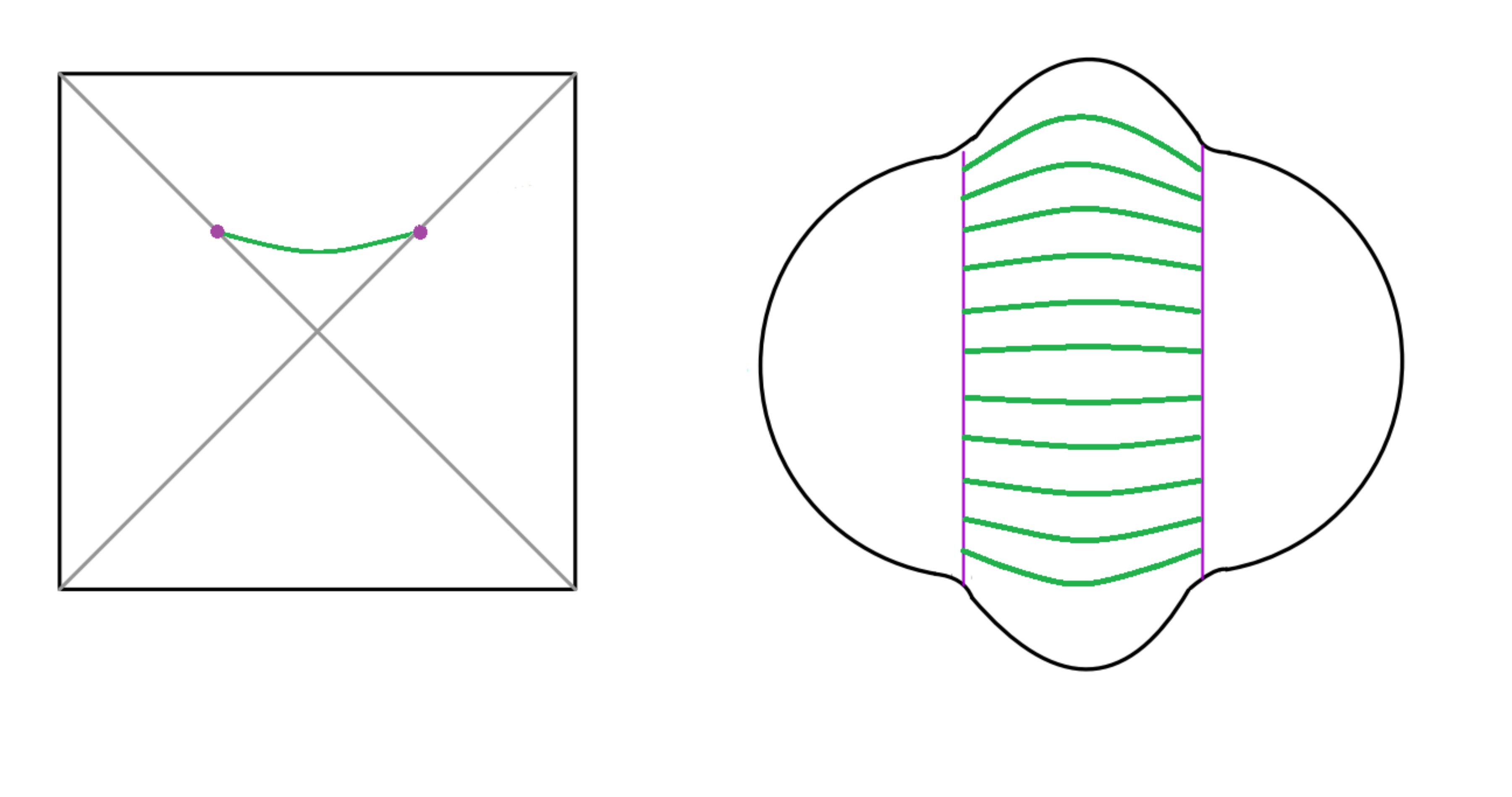}
\caption{A space-like slice containing the green surface between horizons. The right panel shows the \bt \ connecting the horizons. The bottleneck occurs at the horizons.}
\label{pureds}
\end{center}
\end{figure}

Consider pure de Sitter space with the subsets  $\CA$ and $\CB$  taken to be the left (pode) and right (antipode)  horizons. To compute the entanglement entropy between the left and right degrees of freedom we pick a pair of anchor-points  ($D-2$ surfaces)  in the Penrose diagram, lying on the horizons. Then we span the space between them by a $(D-1)$ surface as shown by the green curve on the Penrose diagram in figure \ref{pureds}. 

The right panel of figure \ref{pureds}  shows a space-like slice through the geometry that includes the green curve. 
The bit-threads, emanating from the horizons, are laid out on the $(D-1)$ surface as shown. 

As explained earlier, the reason for the bulge in the embedding diagram is the rapid growth of de Sitter space as one moves into the region between the horizons.
Because of this  growth  the bottleneck limiting the number of bit-threads is  right at the horizon. Therefore entanglement entropy between the left and right horizons is proportional to the area of the horizon according to the usual Gibbons-Hawking formula
\be 
S_{ent} = \frac{A}{4G}.
\ee

Note that in this case the two horizons---left and right---have the same area. The location of the bottleneck is ambiguous but the maximum number of \bts \ is unambiguous.

In general we will want to 
 go a step further along the lines of the HRT \cite{Hubeny:2007xt} formula. That means maximizing the result with respect to the choice of $(D-1)$ surface while  holding the anchoring points fixed \cite{Wall:2012uf}. However since the bottleneck is at the horizons, the result does not depend on the green surface in the left panel of figure \ref{pureds} as long as it is space-like. This gives the horizon area in terms of a maximin surface, whereas naively the (bifurcate) horizon of de Sitter space is a minimax surface which can lead to various pathologies \cite{Shaghoulian:2021cef}. \\

\sc
\section{\S-dS}

\subsection{Fluctuations}

 We will make the following assumptions:  the pode and antipode patches do not interact, but they are entangled. 
At   $t=0,$  the global state is the highly entangled thermofield-double  (TFD) state. And finally, each patch is in thermal equilibrium at the usual de Sitter temperature.

For systems of finite entropy, Boltzmann  fluctuations  can produce configurations that ordinarily would be considered far from equilibrium even though they are part of the thermal ensemble. As an example there is a small probability that a black hole can appear at the center of a static-patch. The probability for such an occurrence is given by
\be 
{\rm Prob} = e^{-\Delta S},
\ee
where $\Delta S$ is the entropy deficit due to the presence of the black hole 
	(see for example \cite{Banks:2006rx}\cite{Susskind:2021dfc}). For black holes much smaller than the de Sitter length it is equivalently the Boltzmann weight,
 \be 
{\rm Prob} = \frac{1}{Z}e^{-\beta M},
\ee
where $\beta $ is the inverse de Sitter temperature, $Z$ is the partition function, and $M$ is the black hole mass.

We may introduce a projection operator $\Pi$  that projects onto states in which such a black hole is present at the pode. The probability to find the black hole may also be expressed as
 \be 
{\rm Prob} = {\Tr} \  \rho  \Pi,
\ee
where $\rho $ is the thermal density matrix of the static-patch.

The projected state with the black hole is given by
\be  
|bh\ra = \Pi |TFD\ra.
\ee
Because of the high degree of entanglement in the TFD state, and the simple form of that entanglement, the presence of a black hole at  the center of  one patch implies the presence of a second black hole at the center of the other patch. Moreover the two black holes will be entangled and are therefore connected by an Einstein-Rosen bridge. 

The projected density matrix of the podal static-patch containing the black hole  is
\be  
\rho_{bh} ={\Tr}_{ap}  \  \Pi |TFD\ra \la TFD|\Pi,
\ee
where ${\Tr}_{ap}$ is the trace over the antipode \dof.

The entanglement entropy between the pode and antipode patches cannot change with time, but one can ask for the \it conditional entanglement entropy,
\rm conditioned on there being a black hole of a given mass at the pode. This is defined by
\be 
S_{cond} = -{\Tr}_p  \ \rho_{bh}  \log{\rho_{bh}},
\ee
where the trace is over the pode degrees of freedom.

It is obvious that $S_{cond} < S$ where $S$ is the equilibrium entropy (as well as the entanglement entropy of the TFD). The difference $\Delta S=(S-S_{cond})$ is the entropy deficit  \cite{Banks:2006rx}\cite{Susskind:2021dfc}     and controls the probability to find the black hole in the pode patch.

\subsection{Entanglement Entropy}

\begin{figure}[H]
\begin{center}
\includegraphics[scale=.4]{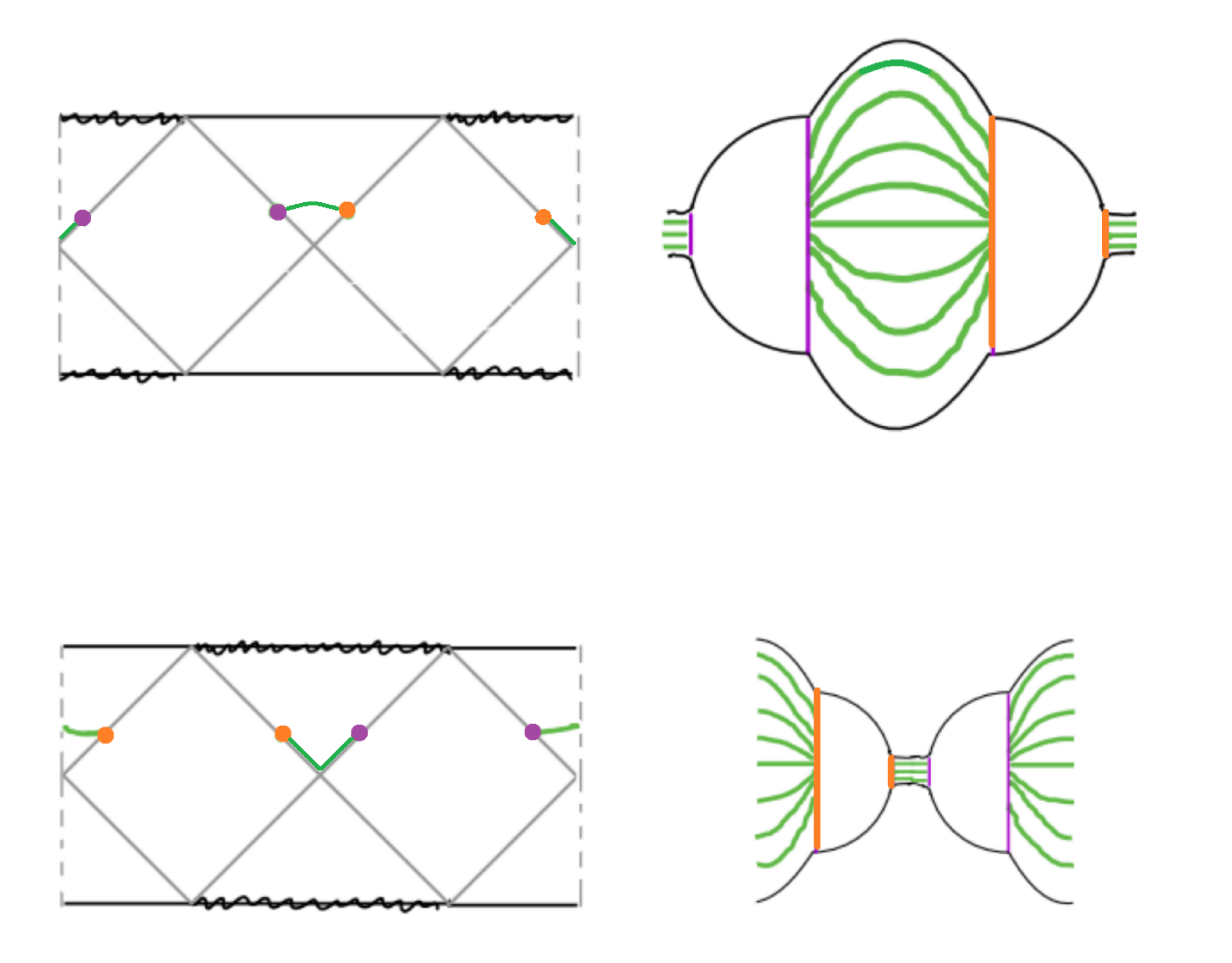}
\caption{The \S-de Sitter geometry shown with \bts \ connecting the two generalized horizons. The diagrams are periodically identified at the dashed vertical lines. The upper and lower panels are identical in content but are identified on different vertical lines.}
\label{dsbh}
\end{center}
\end{figure}

The \S -de Sitter geometry is the gravitational configuration that describes the state $|bh\ra = \Pi |TFD\ra$. In figure \ref{dsbh} we show the periodically identified \S -de Sitter geometry in four different ways. On the left are Penrose diagrams with the identifications made along two different vertical slices. On the right are the corresponding space-like slices.

Both the left and right generalized horizons  have two components--the cosmic horizons and the smaller black hole horizons. Again we draw anchoring points on each side and connect the two sides by space-like surfaces. Each component of the horizon can source bit-threads. The new thing is that bit-threads can thread the wormhole connecting the two black holes. On each component of the horizon the  bit-threads are emitted toward only one side and never cross a horizon.

The bottleneck in this case has two components. There is one bottleneck for the \bts \ passing through the bulge. That bottleneck occurs either at the left or right cosmic horizon--the choice is arbitrary since the two horizons are equal. Moreover, the rule that the area of the bottleneck should be maximized with respect to the surface spanning the two horizons is irrelevant since the bottleneck is independent of that surface.

The other bottleneck is in the wormhole between the black holes. This time the bottleneck is not at the horizon and we do have to maximize the area with respect to the location of the space-like surface \cite{Wall:2012uf}. The reason is that instead of growing, the local $(D-2)$-spheres shrink as one moves into the interior of the black hole. In fact the maximum area occurs when the $(D-1)$-surface dips down to the bifurcate horizon as shown in the lower left panel of figure \ref{dsbh}.

The result is that the (conditional) entanglement entropy of the left and right sides is proportional to the sum of the cosmic horizon area and the area of the  bifurcate horizon of the black hole,
\be 
S_{ent} = \frac{A_{CH} + A_{BH}}{4G}.
\label{shds}
\ee
This is the expected answer (for details see \cite{Susskind:2021dfc}). \\

\subsection{Asymmetric Black Hole}\label{asymbh}

\begin{figure}[H]
\begin{center}
\includegraphics[scale=.4]{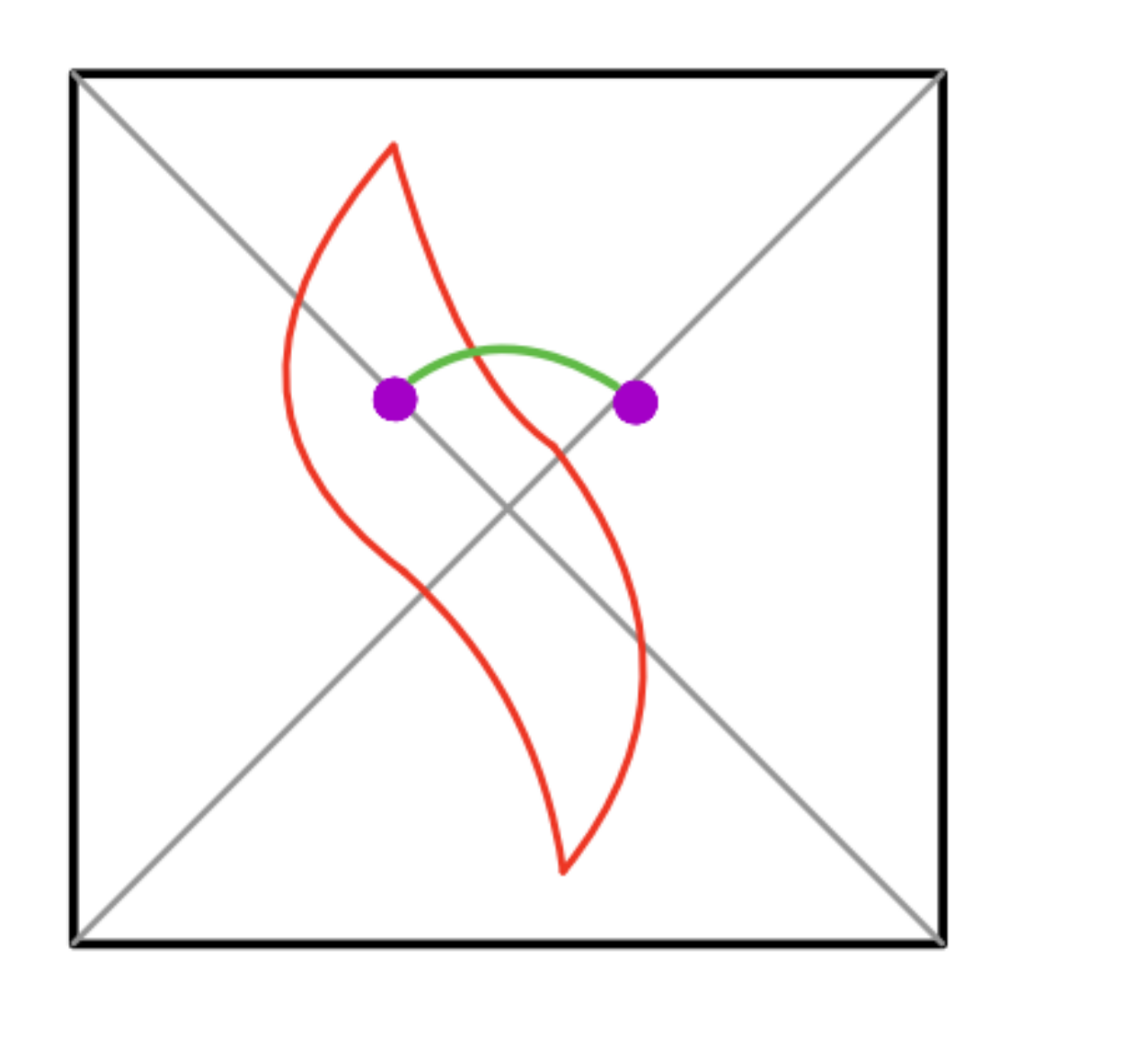}
\caption{Asymmetric black hole pair creation. The red curves represent the world-lines of a pair of black holes which are created and annihilated behind the horizons.  }
\label{pair}
\end{center}
\end{figure}

Another variant of the black hole example goes as follows: We assume that a fluctuation occurs in which an entangled pair of black holes  is produced  and then annihilated  behind the horizon as in  figure \ref{pair}. Each static-patch sees a black hole materialize from the past horizon and disappear into the future horizon. But as figure \ref{pair} illustrates, the process can happen in an asymmetric way relative to the anchoring points. The state is sampled at a time when the left static-patch contains a black hole, but the black hole on the right side has already disappeared behind the horizon. 

\begin{figure}[H]
\begin{center}
\includegraphics[scale=.35]{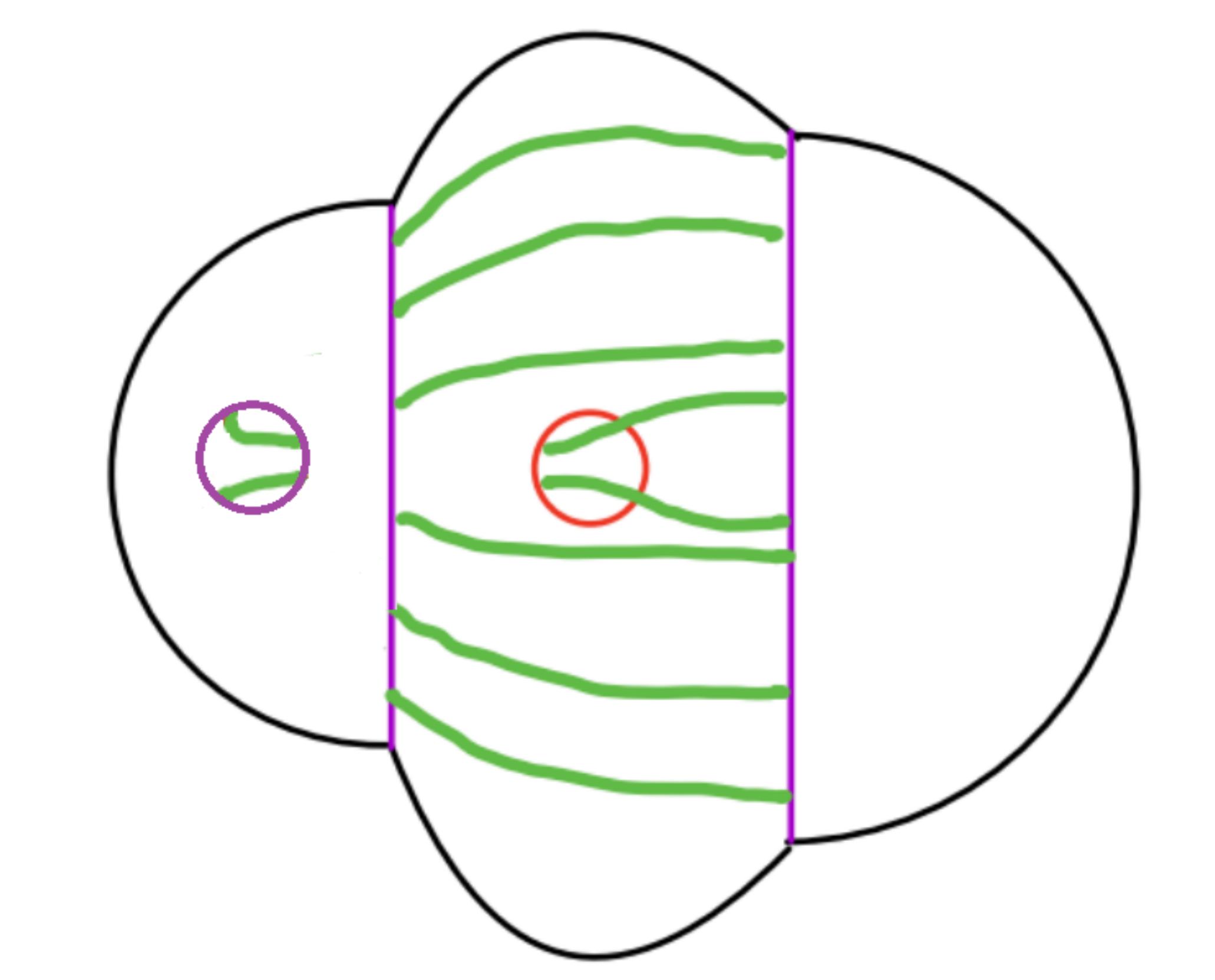}
\caption{Bit-thread diagram for asymmetric black hole pair.}
\label{asym}
\end{center}
\end{figure}

On the green time-slice the black hole horizon on the left is part of the generalized horizon of the left static-patch, but the right static-patch does not contain a black hole. The generalized horizon on the right side consists of a single component. The \bt \ diagram is shown in figure \ref{asym}. Note that the black hole horizon in the left static-patch (shown as purple) is part of the generalized horizon and can therefore be a source of \bts \ while the other end of the wormhole between the horizons (shown as red) is not a true horizon and therefore is not a \bt \ source.
The bottleneck in the diagram is the union of the left-side horizons, namely the left cosmic horizon and the left black hole horizon. Therefore the entanglement entropy is the same as for the symmetric black hole case \eqref{shds}.

Phrased in terms of minimal surfaces, the minimal surface for the right horizon theory is the union of the left cosmic horizon and the left black hole horizon. This means the entanglement wedge of the right horizon theory includes the exterior region between the two cosmic horizons. In \cite{Shaghoulian:2021cef} this was tentatively interpreted as evidence for an interaction between the two horizon theories.

\subsection{Horizon Complementarity}

Imagine following as the right-side black hole passes from the interior of the antipode patch, through the horizon, to the exterior between the horizons.

Comparing figures \ref{dsbh} and \ref{asym} we see something very interesting. In figure \ref{dsbh} the black holes are each in their respective static patches, and each is a source of \bts. But in figure \ref{asym} the antipodal black hole has fallen through the horizon. Once that has occurred
 the corresponding \bt-ends become transferred to the cosmic horizon. This is an explicit example of horizon complementarity--the principle that information which has fallen through the horizon is not lost but becomes  encoded in the horizon degrees of freedom.

\sc
\section{Divided Horizon}
So far in dividing the degrees of freedom into the subsets $\CA, \CB$ connected components of the generalized horizon have not been split. We will now allow such splitting and see where it leads us.

\subsection{Splitting a Horizon}
\begin{figure}[H]
\begin{center}
\includegraphics[scale=.35]{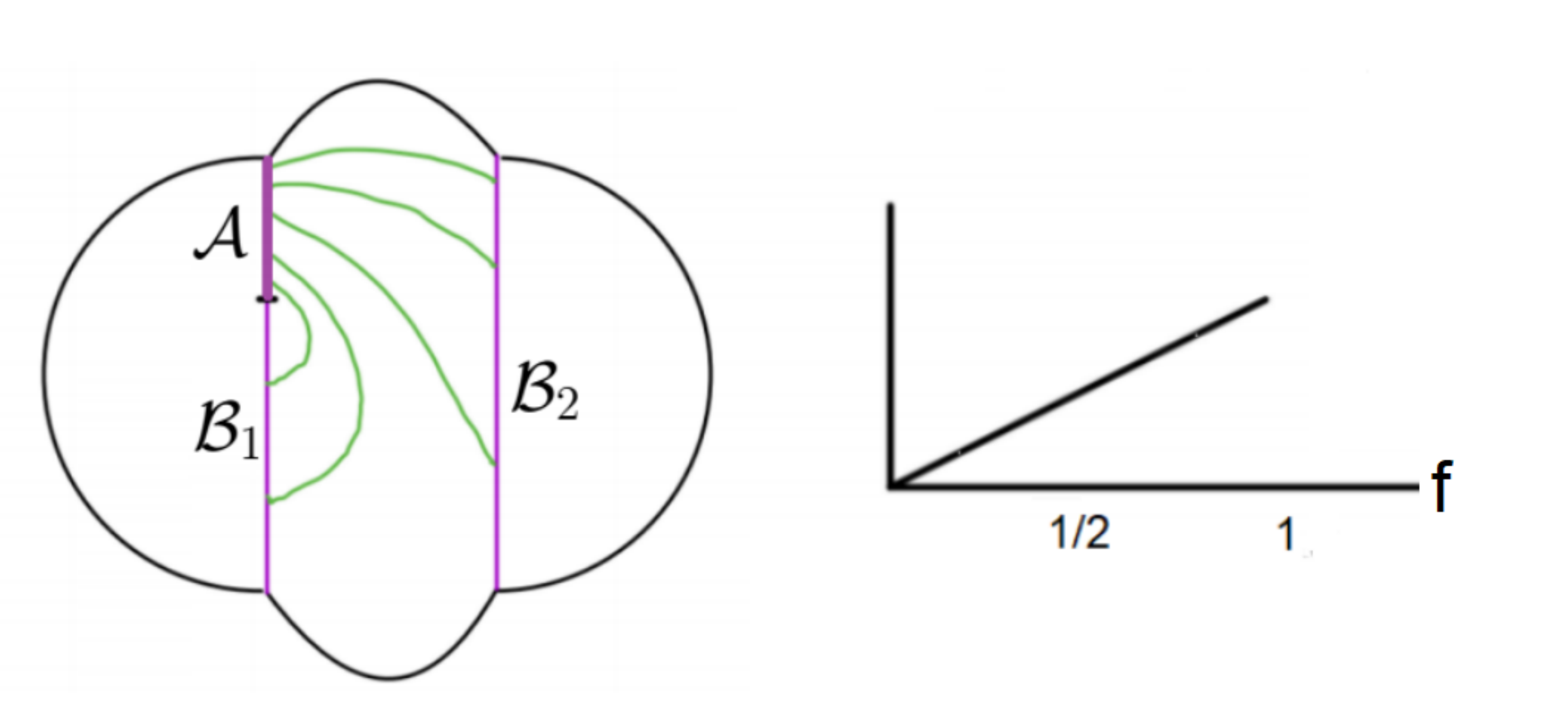}
\caption{The left horizon split into two components and the \bt \ diagram for 
entanglement between $\CA$ and $\CB$.The right panel shows the entanglement entropy as a function of $f.$ }
\label{halfhor1}
\end{center}
\end{figure}

Let's return to pure de Sitter space but consider a different separation of the degrees of freedom into two subsets as suggested  in \cite{Shaghoulian:2021cef}. The division will involve dividing a connected component of a horizon into  parts. For the moment let's suppose that this is done by subdividing the area of the left cosmic horizon into two fractional parts.
A fraction $f$ of the left-horizon \dof \ is identified as the $\CA$ subset. This is shown by a thick purple segment in figure \ref{halfhor1}.

The subset $\CB$ consists of the union of $\CB_1,$ the fraction $(1-f)$ of the left-side horizon; and $\CB_2,$ the full right-side horizon.  Bit threads emanating from $\CA$ can go either to the fraction $\CB_1,$ or to  $\CB_2$. The bottleneck is the portion  of the left-horizon $\CA$.  The entanglement entropy is given by
 \be 
S_{ent} = f \frac{A}{4G}.
\label{obvious}
\ee
where $A$ is area of the full horizon. \\

\subsection{Meaning of Dividing a Horizon}
The holographic description of the boundary of AdS is in terms of a large $N$ quantum field theory. It makes sense to sub-divide the \dof \ of the boundary geometrically into regions and study the entanglement between such regions.

The holographic \dof \ of a de Sitter horizon do not form a spatially local quantum field theory. More likely they are described by a large $N$ $(0+1)$-dimensional system such as matrix quantum mechanics \cite{Banks:2006rx}\cite{Susskind:2021dfc}).  The geometry of the horizon is emergent. The fundamental \dof \ are not attached to geometric points, or even Planck-size cells  on the horizon. 
Geometrically dividing the geometry of a horizon  into sub-regions is not something we ordinarily do.\footnote{One exception is the case of large black holes in AdS when the radius of the horizon is much larger than the AdS scale. On such large scales the horizon behaves like a conventional ``sub-dividable" condensed matter system with a microscopic length scale equal to the AdS scale. The Hartman-Maldacena effect can be thought of in terms of partitioning the horizons of large black holes \cite{Hartman:2013qma}.}

What then do we
 mean by dividing the \dof \ of a horizon  into two subsets? The answer that we propose is simply this: Whatever the fundamental \dof \ are--matrix elements of matrix quantum mechanics or fermions in some dS version of SYK \cite{Susskind:2021esx}--we divide them into subsets.
  For example the $N \times N$ matrix elements of matrix theory could be divided into an $M\times M$ sub-algebra and the  $N^2 - M^2$ remaining elements. In SYK the $N$ fermionic degrees of freedom can be divided into two subsets. In fact the
 SYK theory might provide a testing ground to compute the entropy of a fraction $f$ of the \dof \ and compare it with \eqref{obvious} or \eqref{unobvious}.

\subsection{A General Argument} \label{general}
Equation \eqref{obvious} is easy to understand given certain simplifying assumptions. We will lay them out  and comment on them here and more fully in section \ref{TL}.
Let's model the two cosmic horizons as identical systems, each of $N$ qubits. 
The left system is partitioned into two subsystems of $fN$
and $(1-f)N$ qubits. We assume that $f$  is large enough that the subsystem of $fN$ qubits is macroscopic.
Let us also assume the left and right subsystems are non-interacting but are entangled in the TFD state.

In dimensionless Rindler units the temperature characterizing the TFD state is $T=1/2\pi.$ It is neither extremely large nor extremely small.
If the number of qubits in each horizon is very large the horizons will naively  satisfy the criteria for the conventional thermodynamic limit (TL) to apply. 

In the TFD state the entropy $S$ is  the fine-grained entanglement entropy, and it is also the coarse-grained thermal entropy of either of the horizons. Now consider the fractional subsystem of size $fN$ (call it $f$ for simplicity).  By standard arguments based on the thermodynamic limit, its coarse-grained entropy is $fS.$ Because fine-grained entropy is always less-than-or-equal to coarse-grained entropy it follows that the fine-grained entropy satisfies,
\be 
S_{ent}\leq fS.
\label{SleqfS}
\ee
In general this inequality will be close to saturated if $fN$ is large enough that the surface-to-volume ratio for the subsystem $f$  is negligible. Thus, given that the thermodynamic limit applies, \eqref{obvious} follows from fairly general arguments. \\

The problem with this argument is that there are good
 reasons to doubt that the thermodynamic limit and the consequent simple additivity of entropy, applies. The standard TL depends on spatial locality such as the kind that quantum field theories and ordinary lattice theories enjoy. Holographic theories are not of this type. Generally they are ``all-to-all" \kl \ theories which are much more densely and non-locally coupled. There is reason to think that de Sitter space is described by even more non-local ``hyperfast" dynamics. Theories with this kind of dynamics generally do not have a conventional thermodynamic limit. The bilayer proposal that we come to next manifests non-thermodynamically limiting behavior.

\sc
\section{The Bilayer Proposal }

Reference \cite{Shaghoulian:2021cef}  suggested a ``bilayer" modification of the monolayer proposal.  
 The arguments in  \cite{Shaghoulian:2021cef}  are based on the minimal area principle of Ryu and Takayanagi, while this paper uses the \bt \ version of Freedman and Headrick. This is a trivial difference: as Freedman and Headrick explain,  the max flow-min cut theorem insures that the  minimal area principle and the maximum  \bt \  principle are equivalent. The \bt \ method is easily adaptable to the bilayer version and easy to visualize. The substantive differences between the proposals are described below in \bt \ language:\\

%The advantage of bilayer construction is that it allows us to focus only on the %largest components of the generalized horizons (the cosmic horizons) as %the sources of \bts. Expressed in \bt \ terms, the mono and bilayer theories %are different in two important ways:

\begin{enumerate}
\item
In the monolayer theory horizons are single layers whose area density of \bt \ emanations is bounded by 1/4G.

 By contrast in the bilayer theory each cosmic horizon is a double-layered surface. Each of the layers can emit \bts \ as shown in figure \ref{twolayer}. For \it each \rm layer  the area-density of emitted \bts  \ is bounded by $1/4G.$ 
\begin{figure}[H]
\begin{center}
\includegraphics[scale=.5]{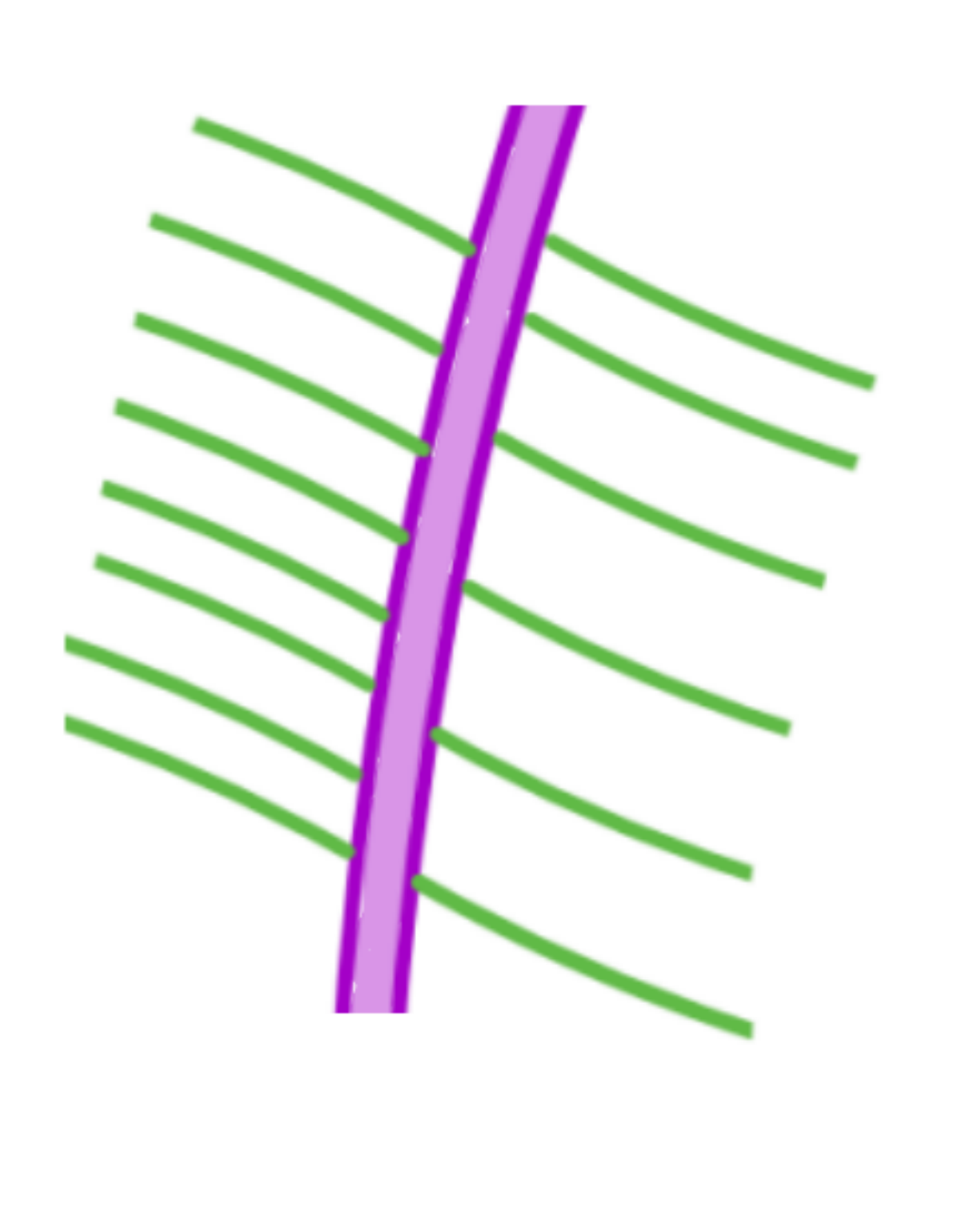}
\caption{In the bilayer proposal a horizon is composed of two layers which can independently source \bts. The area density of \bts \ on each layer is bounded by $1/4G.$}
\label{twolayer}
\end{center}
\end{figure}
 As before \bts \ may not cross a horizon. 

\item In the monolayer theory all horizons must source \bts. In the bilayer theory we will count only the largest  components on each side (the cosmic horizons) as sources of \bts.\footnote{We could consider all horizons as sources of bit-threads in the bilayer theory, but this will not make a difference for the cases we consider in this paper.}

\end{enumerate}

\subsection{Pure De Sitter}
\begin{figure}[H]
\begin{center}
\includegraphics[scale=.4]{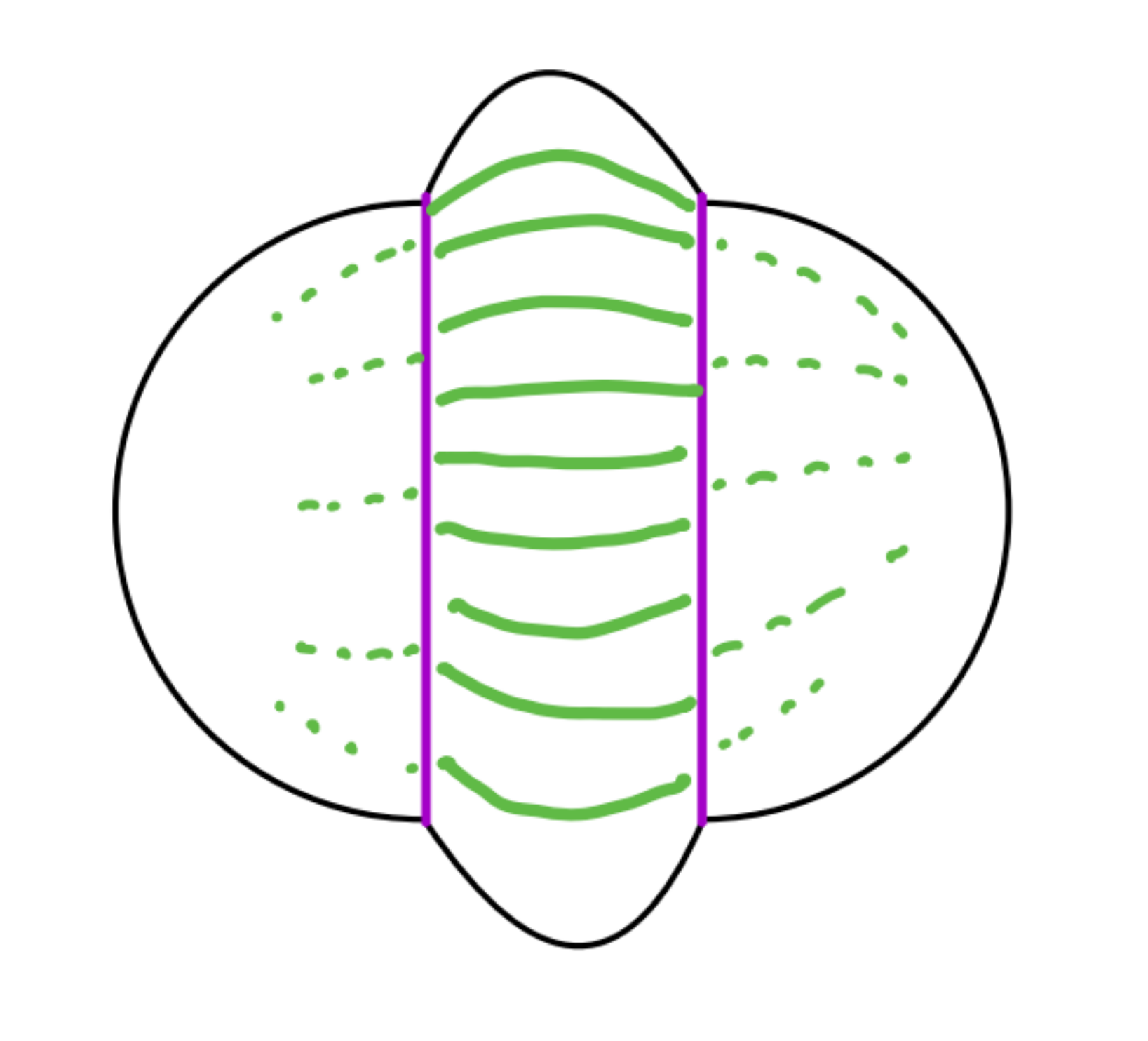}
\caption{Bit-threads emitted toward the pode and antipode encounter a bottleneck of zero area.}
\label{twopure}
\end{center}
\end{figure}

Let's go through the various examples to see how the bilayer model works. First the pure de Sitter example in which the two horizons comprise the subsets $\CA$ and $\CB.$ Going back to figure 
\ref{pureds} the only modification to the \bt \ diagram would be additional \bts \ emitted toward the pode and antipode as indicated by dashed lines in figure \ref{twopure}.
But there is no route for those \bts \ emitted from one horizon to get to the horizon on the opposite side. They encounter an absolute bottleneck of zero area. Thus the bilayer model adds nothing to the monolayer theory and gives the same result. \\

\subsection{\S-De Sitter}
\begin{figure}[H]
\begin{center}
\includegraphics[scale=.5]{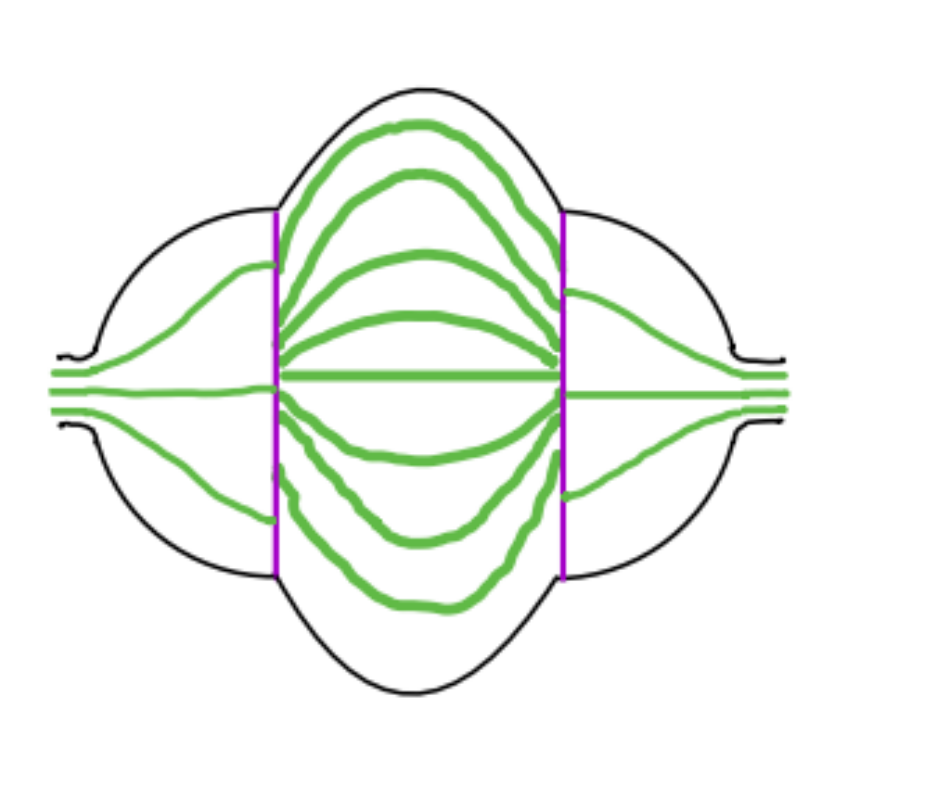}
\caption{ Bit-threads emitted from the horizons can thread the Einstein-Rosen bridge connecting the black holes.}
\label{twoside}
\end{center}
\end{figure}

Next consider the \S-de Sitter geometry in in the left panel of figure \ref{dsbh}. The new features are that the black hole horizons do not emit \bts,  but the cosmic horizons can emit \bts \ to both sides. Figure \ref{twoside} shows the bit-thread diagram that replaces the right panel of figure \ref{dsbh}.
%
%It is clear that the effects of the two new features cancel one another. 
The black hole horizons no longer emit \bts; instead they act as a  bottleneck for \bts \ that thread the wormhole. The \bts \  threading the wormhole are emitted from the second layer of the cosmic horizons instead of the black hole horizons.  Again, the net result is \eqref{shds} and is unmodified from the corresponding monolayer result.\\

\subsection{Asymmetric Black Hole}
\begin{figure}[H]
\begin{center}
\includegraphics[scale=.4]{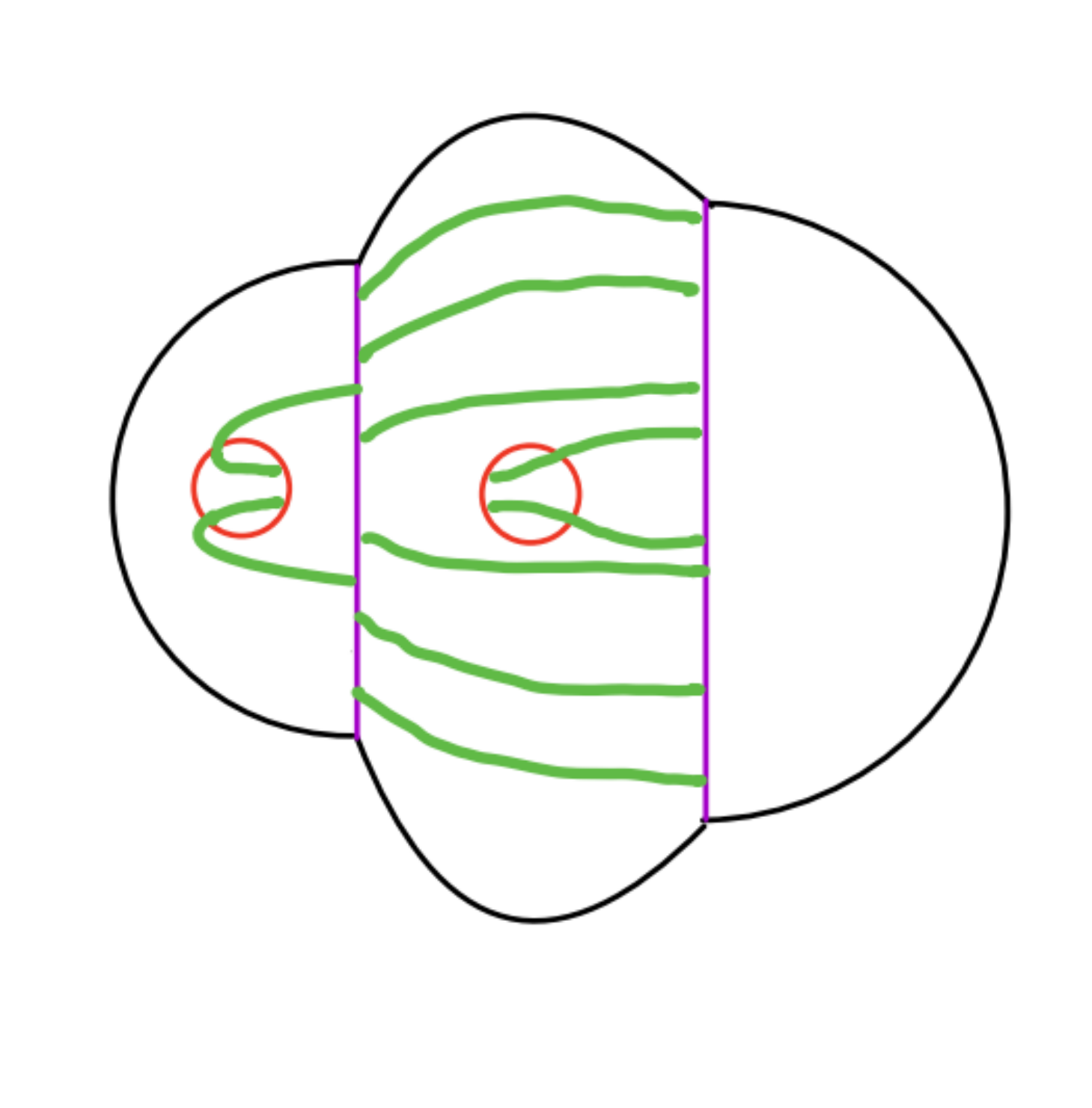}
\caption{Bit-thread diagram for asymmetric black hole pair-creation in the bilayer theory.}
\label{twosd}
\end{center}
\end{figure}

The same is true for the asymmetric black hole example in figure \ref{asym}. The diagram for the bilayer theory is shown in figure \ref{twosd}.
\bn
Again it is easy to see that the bilayer theory gives the same answer as the monolayer theory. \\

\sc 
 \section{A Disagreement}
 The mechanism by which the monolayer and bilayer theories agree is so simple that one might think it is both general and trivial. But the next example, described in \cite{Shaghoulian:2021cef}, will show that this is not the case.

\subsection{The Split Horizon in the Bilayer Theory}
\begin{figure}[H]
\begin{center}
\includegraphics[scale=.45]{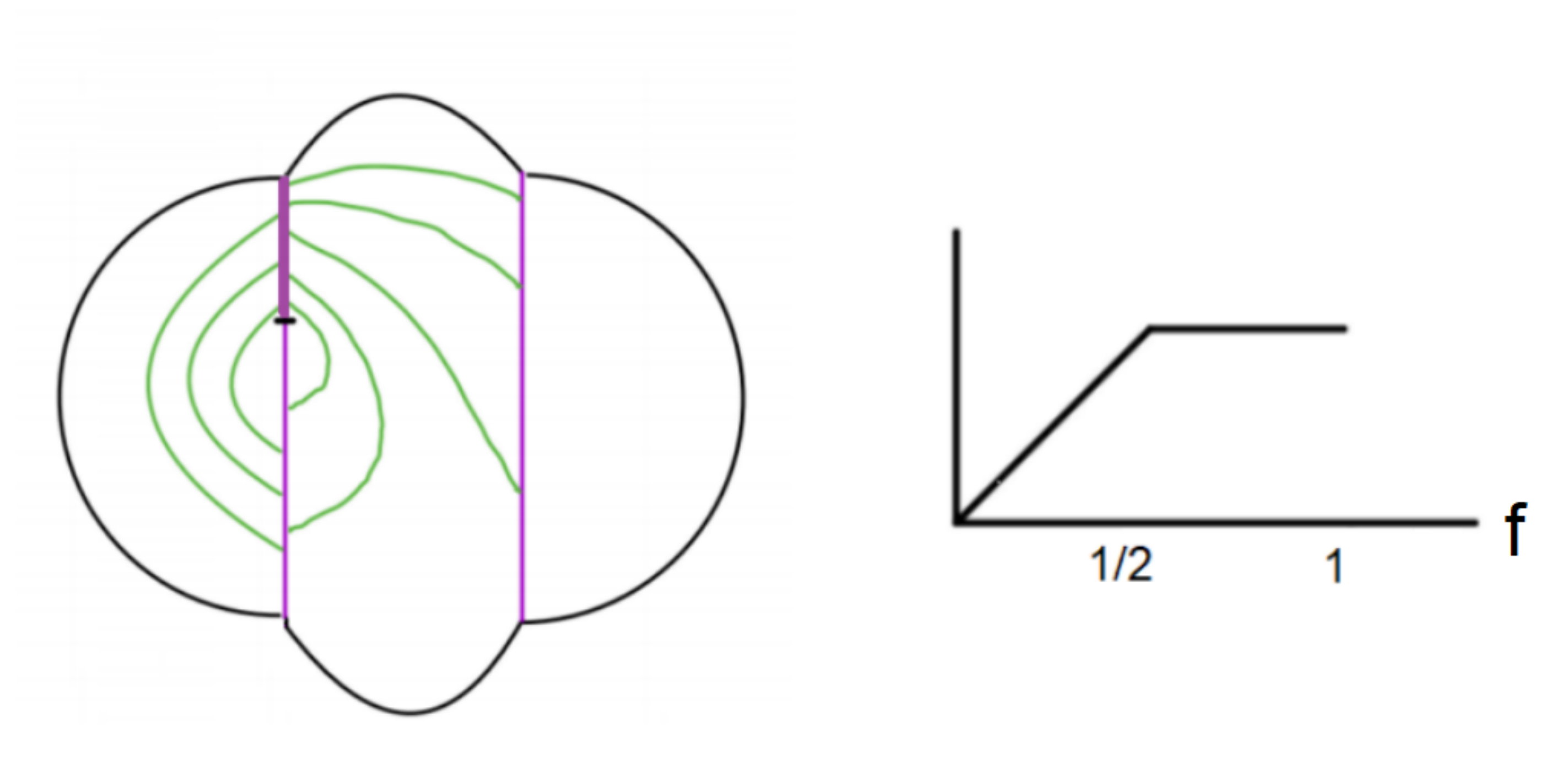}
\caption{Bit-thread diagram for a split horizon in the bilayer theory. The right panel shows the entanglement entropy as a function of $f$.}
\label{halfhor2}
\end{center}
\end{figure}

 Consider the entanglement between $\CA$ and $\CB_1 \cup \CB_2$  illustrated in figure \ref{halfhor1}. In the bilayer theory both cosmic horizons may emit \bts \ in either direction, but those emitted to the right from $\CB_2$ have not place to go. So they don't contribute to the entanglement. The new \bts \ that do contribute are shown in figure \ref{halfhor2}.
\bn
They are the \bts \ that  emanate to the left from the left-side horizon. It is clear that their contribution is non-zero, thus breaking the equivalence of the monolayer and bilayer formulations.

When $f<1/2$ the bottlenecks for both sets of \bts \ are the segment $\CA.$ the result is that the entanglement grows linearly with $f$, but \it twice as fast \rm as for the monolayer case. But once $f\geq 1/2$ the bottleneck for the ``exterior" \bts \ is still $\CA,$ but the bottleneck for the interior \bts \ is $\CB_1.$ It follows that the number of \bts \ is constant in this range of $f.$ The predicted  entanglement entropy  shown in the right panel of figure \ref{halfhor2} has the form \cite{Shaghoulian:2021cef},
\bea
S&=&2f \frac{A}{4G}  \ \ \ \ \ \ \ \ (f<1/2) \cr \cr
S\eq  \frac{A}{4G}    \ \ \ \ \ \ \ \  (f>1/2)
\label{unobvious}
\eea

\subsection{A Discrepancy}
Comparing  \eqref{obvious} and \eqref{unobvious}, we see that the two theories (monolayer and bilayer) are not equivalent.
As explained in section \ref{general} the linear behavior of the entanglement entropy over the full range $0\leq f\leq 1$ 
is expected from  the assumption of the thermodynamic limit. The behavior in \eqref{unobvious}  appears to violate those assumptions. For example, at the point where $f=1/2$ the subsystem $\CA$ already has as much entropy as the full left-side horizon. From the viewpoint of conventional many-body theory this  behavior seems bizarre, but we will argue that it is not unreasonable for the kinds of holographic theories that  describe horizons. \\

\subsection{The Thermodynamic Limit}\label{TL}
In the thermodynamic limit (TL) certain quantities such as energy and entropy are extensive. That means they are additive. Linear growth with sub-system size, as described in \eqref{obvious} is natural in the TL. 
What are the criteria for the TL to apply? One important one is spatial locality. Related to this is that the subsystems be large enough, and the surface-to-volume ratio be small enough, so that surface effects can be ignored. For example a lattice system in $D$ dimensions has the property that the surface-to-volume ratio for a subsystem of $M$ lattice points would scale like 
\be  
 \frac{\rm{surface}}{\rm{volume}}  = M^{-\frac{1}{D}}.
 \label{stov}
\ee
 By making the subsystem large enough, but still smaller than the entire system, the surface-to-volume ratio can be made arbitrarily small.

However, the kind of holographic systems that represent horizons are not spatially local.
For black holes they are \kl \ all-to-all  fast scramblers which in many
 ways behave as if $ D \to \infty $ in \eqref{stov}.
Worse still, for de Sitter space they may be even more non-local  hyperfast scramblers \cite{Susskind:2021esx} .
It would be surprising if such systems satisfy the conventional behavior associated with the TL.

There is one situation where the TL works, even for very non-local Hamiltonians and that is the infinite temperature limit. The density matrix in that case is the maximally mixed density matrix, independent of the Hamiltonian. If de Sitter space can be modeled as an 
infinite temperature limit then the TL may apply. But at finite temperature it would be unlikely for \eqref{obvious} to apply.

At the moment we are unable to rule out the monolayer behavior \eqref{obvious}	or the  bilayer behavior \eqref{unobvious} for de Sitter space horizons. 

\subsection{Models}\label{sec:models}
Before concluding this section we will present some evidence that \eqref{unobvious} is a consistent behavior. The basic point is that thermodynamics in large-$N$ systems is distinct from thermodynamics in large-volume limits, even though large $N$ is sometimes referred to as a thermodynamic limit. For example, for holographic systems dual to AdS spacetime, the canonical and microcanonical ensembles are inequivalent, due to small black holes which are microcanonically stable but have negative specific heat. In the ordinary thermodynamic limit these ensembles are instead equivalent. Coupled SYK systems have a similar inequivalence, due to the hot wormhole phase of \cite{Maldacena:2018lmt}.\footnote{In many holographic systems the existence of a higher-form symmetry trivializes many -- but not all -- finite-size effects and helps mimic a large-volume thermodynamic limit \cite{Shaghoulian:2016xbx}\cite{Shaghoulian:2020omr} through the Eguchi-Kawai mechanism \cite{Eguchi:1982nm}. For example, the lack of finite-volume corrections to the thermodynamic entropy above the Hawking-Page transition \cite{Hartman:2014oaa}\cite{Belin:2016yll} can be explained this way.}

\begin{figure}[H]
\begin{center}
\includegraphics[scale=.35]{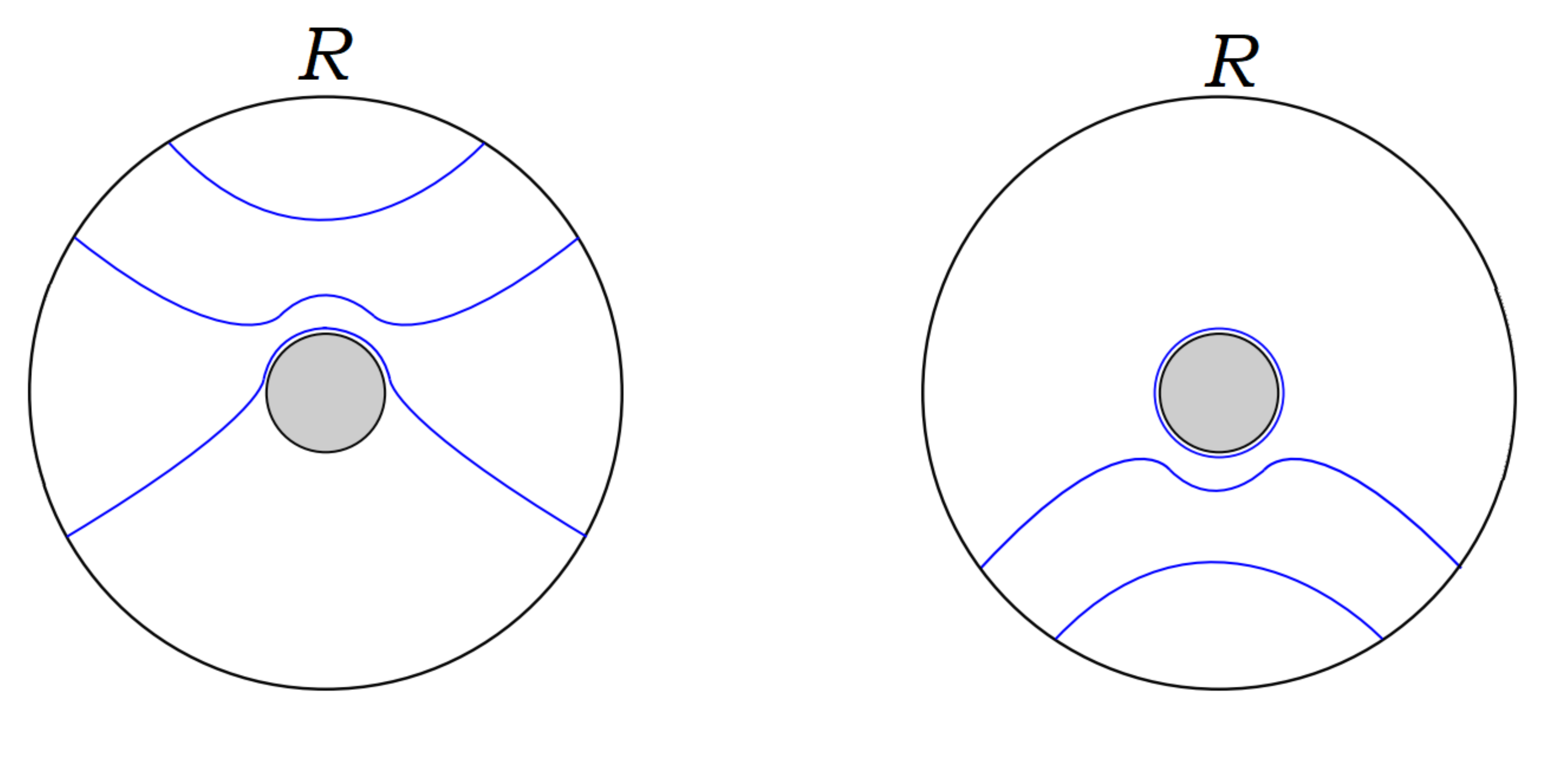}
\caption{We consider the minimal surface for a region $R$ on one side of the thermofield double in AdS$_3$. There is a transition from the left diagram to the right, where the minimal surface becomes disconnected and has a component which coincides with the event horizon of the black hole. This happens at an $O(1)$ fraction of the boundary system size.}
\label{bhsurfaces}
\end{center}
\end{figure}

More explicitly, we can see finite-size effects which are absent in a traditional thermodynamic limit. Consider the thermofield double state of a holographic CFT, dual to two entangled black holes. We consider a temperature of the same order as the size of the spatial manifold. We can compute the entropy of a subregion of one of the two CFTs using the Ryu-Takayanagi formula. The case of AdS$_3$ is shown in figure \ref{bhsurfaces}. For small region sizes the minimal surface stays near the boundary and is weakly affected by the presence of the black hole. Small regions probe the ultraviolet so this is expected. As the region size grows the minimal surface feels the presence of the black hole, running near and along a portion of it for large enough system sizes. At an $O(1)$ fraction of the system size, there is a phase transition in the minimal surface, and it disconnects. There are now two pieces, one which coincides precisely with the bifurcate horizon, and one which is anchored to the AdS boundary. As we grow the region further the piece anchored to the AdS boundary shrinks to zero size. This phase transition at an $O(1)$ fraction of the system size is similar to what we saw in the double-layered model, and like in that situation it is due to vertical entanglement in the system (i.e. entanglement within one CFT in the thermofield double, as opposed to horizontal entanglement which is between the two CFTs). At infinite temperature the entanglement is entirely horizontal and the entropy curve becomes strictly linear: this recovers the predictions of the thermodynamic limit, since infinite temperature in a CFT is the same as infinite volume. At zero temperature the entanglement is entirely vertical, and we recover the answer in vacuum AdS where the entropy grows and then shrinks down to zero. The case we studied above at finite temperature is an intermediate situation.

This case is slightly complicated by the presence of ultraviolet degrees of freedom which pollute the entanglement we hope to characterize. This is because the holographic dual is a continuum quantum field theory. We expect the dual for de Sitter to instead be a finite system. This brings us to our next example, which has a finite Hilbert space. In fact, it will be just two qubits, called $\sigma$ and $\tau$. The Hamiltonian is the antiferromagnetic coupling
\be
H = J\sigma \cdot \tau,
\label{Heis} 
\ee
where $J$ is an energy scale.

We will calculate the entropy of $0$ qubits, $1$ qubit, and $2$ qubits as functions of temperature. Let us begin with $T=\infty.$ One finds,
\bea 
S_0 \eq 0 \cr 
S_1 \eq \log{2} \cr
S_2 \eq 2\log{2}.
\label{infT}
\eea
\begin{figure}[H]
\begin{center}
\includegraphics[scale=.5]{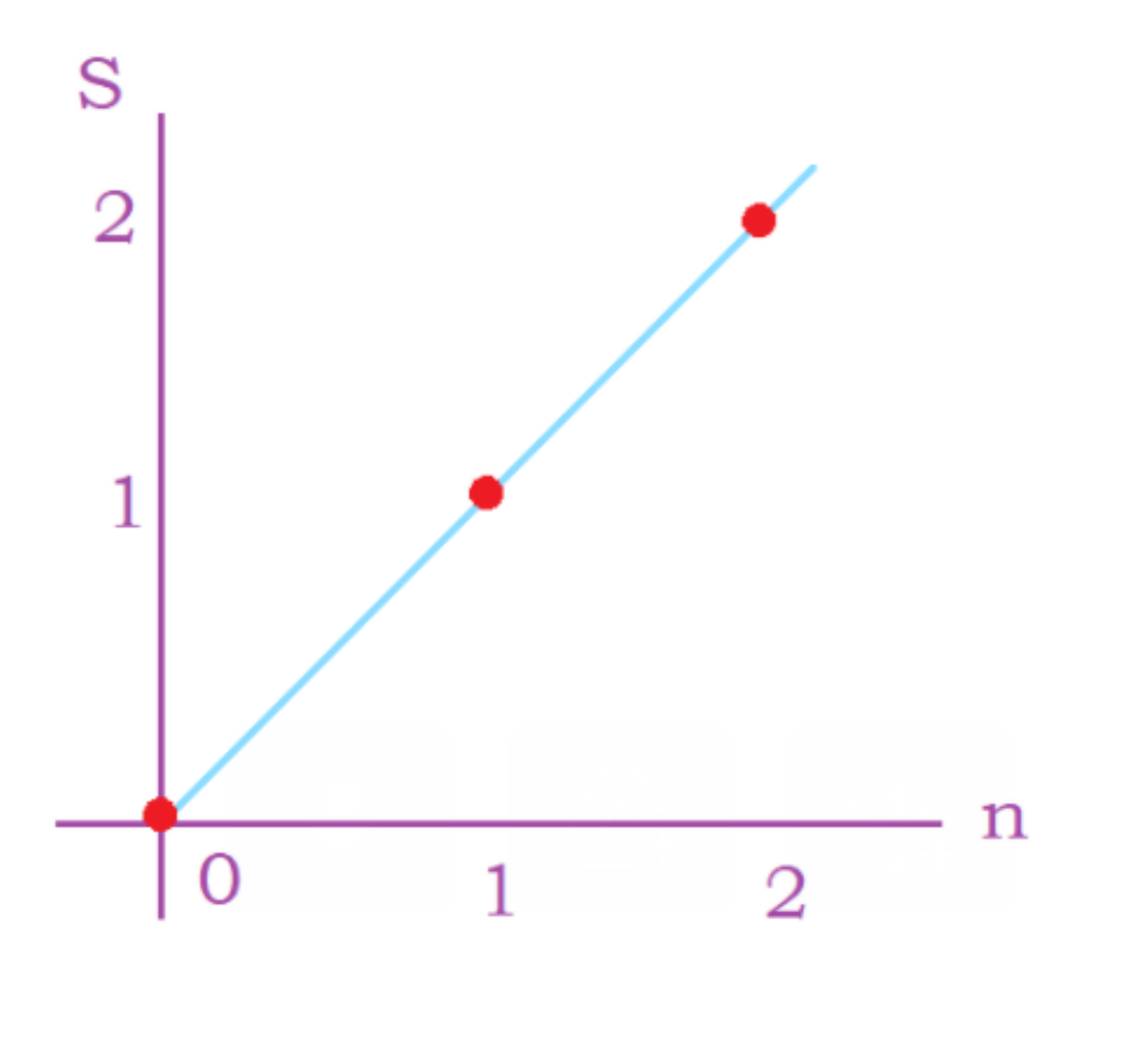}
\caption{}
\label{heis1}
\end{center}
\end{figure}
This is illustrated graphically in figure \ref{heis1}. This is analogous to the behavior in \eqref{obvious}. \\

\begin{figure}[H]
\begin{center}
\includegraphics[scale=.5]{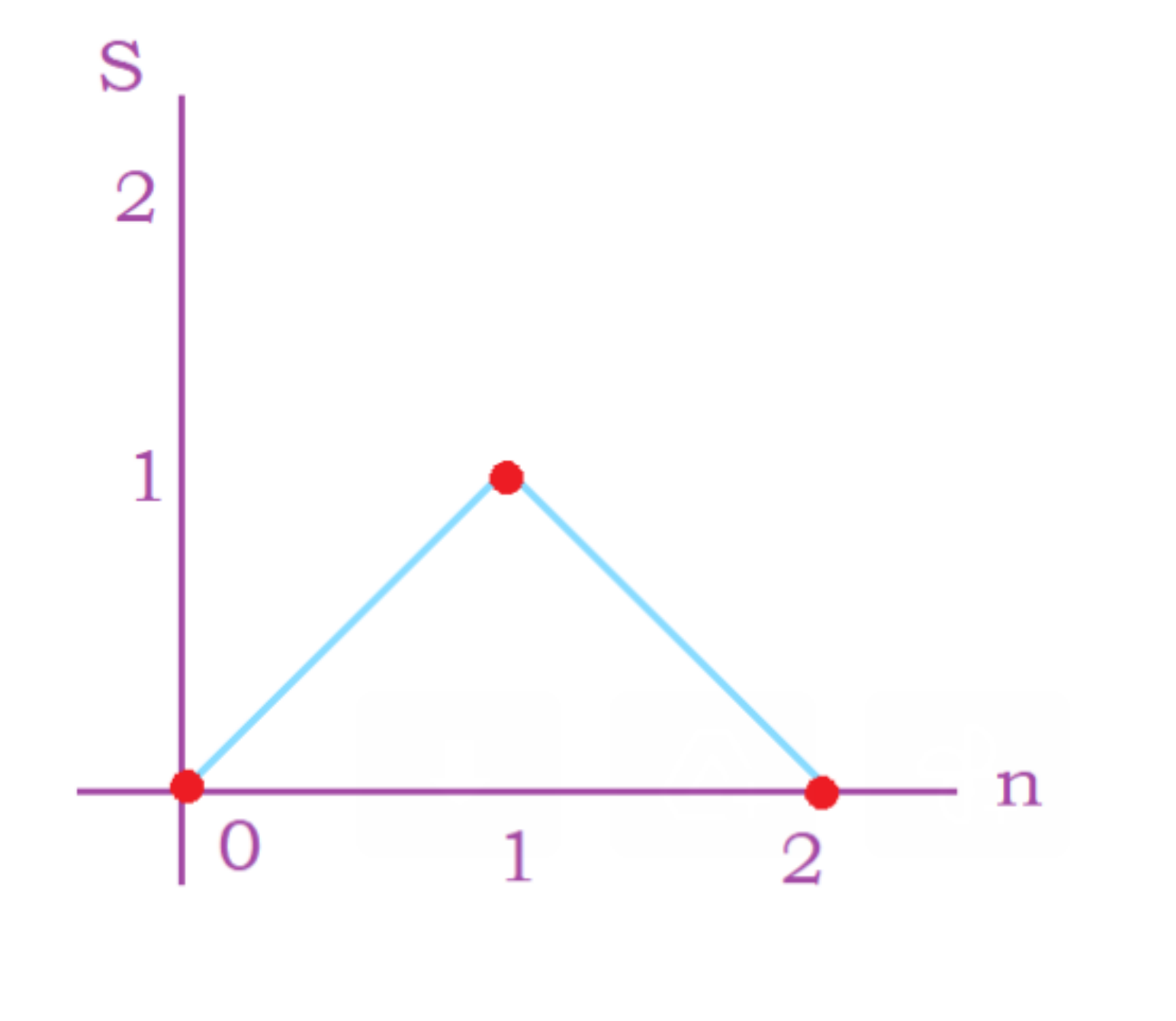}
\caption{}
\label{heis2}
\end{center}
\end{figure}

Now consider $T=0$. In this case the thermal density matrix descibes the pure ground state. The entropy of the zero-qubit subsystem is of course still zero. It is also easy to see that the entropy of the one-qubit 
subsystem is again $\log{2}.$  But for zero temperature the entropy of the two-qubit system is zero, the reason being that the ground state is pure. This is illustrated in figure \ref{heis2}.

At finite non-zero temperature the one-qubit subsystem continues to have entropy $\log{2}$ for all temperatures. This follows from the $SU(2)$ invariance of the thermal state which requires the one-qubit density matrix to be invariant. The only invariant one-qubit density matrix is the maximally mixed density matrix.

\begin{figure}[H]
\begin{center}
\includegraphics[scale=.5]{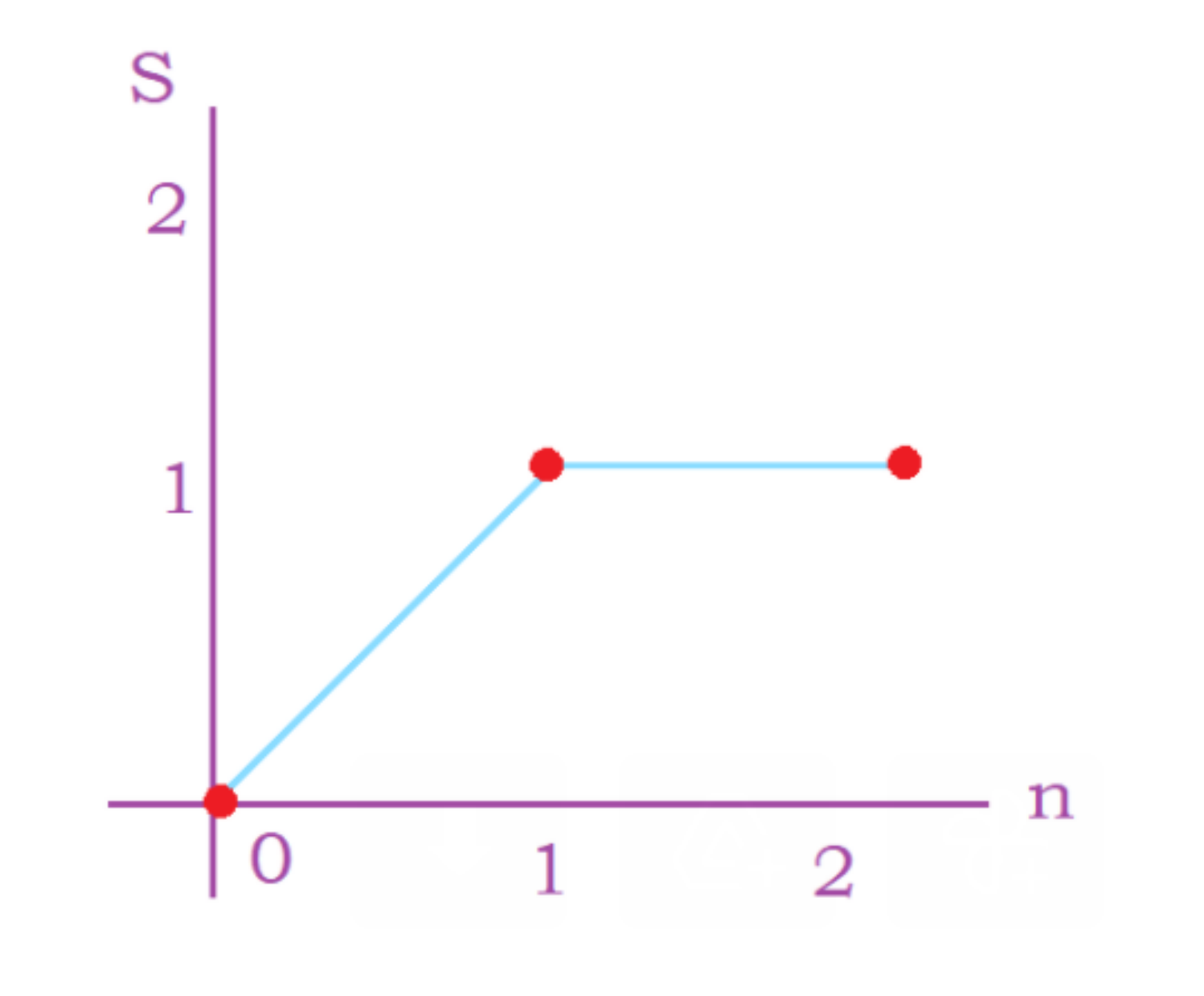}
\caption{}
\label{heis3}
\end{center}
\end{figure}

The two-qubit entropy is not constrained by $SU(2)$ and smoothly interpolates between the zero and infinite temperature cases. At some temperature it will be $\log{2}$ and the graph will be as in figure \ref{heis3}.
\bn
Figure \ref{heis3} is of course the analog of the right panel of figure \ref{halfhor2}. The temperature at which this occurs is easily calculated and is given by $$T = 1.56 J.$$

Before completing this section we would like to put forward a conjecture that generalizes the behavior in figure \ref{heis3}. An interesting feature of the curve is that although the total entropy of the full system (the $n=2$ point) is less that the corresponding entropy in the infinite temperature limit, the curve in the region $n\leq 1$ follows the infinite temperature limit. We suspect that this may well be more general for \kl \ all-to-all systems. If it is true then it would imply that small subsystems behave as if the system had a thermodynamic limit characteristic of infinite temperature and only when the fraction $f$ exceeds $1/2$ does the behavior deviate from the TL.\\

At this point the skeptic will say, you've shown that the behavior \eqref{unobvious} is consistent, but only for a very small system. Why should I believe it can happen for large systems? 

The true believer will answer that  all-to-all systems always behave like small systems: their surface-to-volume ratio is always large.

\sc
\section{Conclusion}
For cases in which connected components of a horizon are not subdivided, both the monolayer and bilayer generalizations of the RT formula  proposed in   \cite{Susskind:2021dfc}\cite{Susskind:2021esx}\cite{Shaghoulian:2021cef} 
give  good accounts of entanglement entropy in de Sitter space, for all the cases we looked at (and some that we didn't) in this paper. The results are in accord with expectations about de Sitter space and standard ideas about quantum entanglement. 

A new example was given involving asymmetric black holes in which on one side the black hole is in the static pode-patch but on the other side the entangled black hole has fallen through the horizon of the antipode-patch. This is particularly interesting because it illustrates that information in a system which has fallen through the horizon is actually encoded on the horizon  (see figure \ref{asym}).

But in cases where horizons are split the situation is not so clear.
The example studied in this paper leads to a discrepancy between the monolayer and bilayer theories and may allow us to decide between them. The monolayer  result  agrees with  what one would expect, assuming an ordinary  thermodynamic limit \eqref{obvious}, while the bilayer theory  does not.  But as we explained, there are strong reasons to doubt that the  holographic systems  which  describe horizons, satisfy the criteria for a  conventional TL. At the moment this leaves us without a definitive conclusion.

One way to try to resolve this issue would be to study the dependence of entropy on subsystem-size in various holographic models such as SYK and the double-scaled limit of SYK.
We hope to come back to this in future work.

\subsection*{Note Added}
A paper which studies entropy as a function of subsystem size \cite{Zhang:2020kia} was brought to our attention. A number of features stand out. In general, for finite temperature, the behavior for small $f$ follows the infinite temperature behavior until it breaks away at some $f\sim 1/2.$  The full entropy at $f=1$ is a temperature-dependent fraction of the infinite temperature value, which for some value of $T$ is $1/2.$ The main difference between the calculated curve of \cite{Zhang:2020kia} and the bilayer curve in figure \ref{halfhor2} is that the calculated curve is smooth with no sharp kink. It behaves similarly to the entanglement entropy of thermal holographic CFTs (upon vacuum subtraction) considered at the beginning of Section \ref{sec:models}. Nevertheless we think this is supporting evidence for bilayer behavior in SYK.

\section*{Acknowledgements}

We thank Ying Zhao for helpful discussions and for calling our attention to reference \cite{Zhang:2020kia}. ES is supported by the Simons Foundation It from Qubit collaboration (385592) and the DOE QuantISED grant DESC0020360. LS is supported in part by NSF grant PHY-1720397.

\end{document}